\documentclass[aps,preprint,prd,showpacs,nofootinbib]{revtex4-1}

\usepackage{epsf}

\usepackage{bm}
\expandafter\let\csname equation*\endcsname\relax
\expandafter\let\csname endequation*\endcsname\relax
\usepackage{amsmath,amssymb}
\usepackage{amsfonts}
\usepackage{subfigure}
\usepackage{graphicx}
\usepackage{natbib}

\def\be{\begin{equation}}
\def\ee{\end{equation}}
\def\bea{\begin{eqnarray}}
\def\eea{\end{eqnarray}}

\newcommand{\bmat}{\left(\begin{array}}
\newcommand{\emat}{\end{array}\right)}

\newcommand{\bit}{\begin{itemize}}
\newcommand{\eit}{\end{itemize}}
\newcommand{\bnu}{\begin{enumerate}}
\newcommand{\enu}{\end{enumerate}}

\newcommand{\ba}{\begin{align}}

\begin{document}

\title{Jeans analysis of Bok globules in $f(R)$ gravity}

\author{Jaakko Vainio and Iiro Vilja}
\affiliation{Department of Physics and Astronomy, Turku Center for Quantum Physics, University of Turku, FIN-20014, Finland}
\email{jaakko.vainio@utu.fi, vilja@utu.fi}
\pacs{04.50.Kd,04.25.Nx,98.80.Es}

\begin{abstract}
We examine the effects of $f(R)$ gravity on Jeans analysis of collapsing dust clouds. We provide a method for testing modified gravity models by their effects on star formation as the presence of $f(R)$ gravity is found to modify the limit for collapse. In this analysis we add perturbations to a de Sitter background. As the standard Einstein-Hilbert Lagrangian is modified, new types of dynamics emerge. Depending on the characteristics of a chosen $f(R)$ model, the appearance of new limits is possible. The physicality of these limits is further examined. We find the asymptotic Jeans masses for $f(R)$ theories compared to standard Jeans mass. Through this ratio, the effects of the $f(R)$ modified Jeans mass for viable theories are examined in molecular clouds. Bok globules have a mass range comparable to Jeans masses in question and are therefore used for comparing different $f(R)$ models. Viable theories are found to assist in star formation. 
\keywords{modified gravity \and f(R) theories \and Bok globules \and Jeans analysis}
\end{abstract}

\maketitle

\section{Introduction}
The standard cosmological model, also called the concordance model \cite{Wang:1999fa}, is based on general relativity (GR) combined with cold dark matter and the cosmological constant. It explains
nicely almost all observational data, in particular the accelerated expansion of the Universe \cite{Perlmutter1999,riess98}. However, the nature and smallness of the cosmological constant 
is highly problematic as there is no natural way to generate such an extreme parameter \cite{Weinberg:1988cp}. Therefore, there are various competing models including new forms of matter \cite{Bilic200217,Bento:2002ps}, inhomogeneous cosmologies \cite{Bolejko:2011jc} and modified gravitation theories \cite{Clifton:2011jh}. The alternative explanations are all constrained by local experiments showing that general relativity works well in stellar system and galactic scales. Therefore, the cause of the accelerated expansion must be restricted to large scales. 

One widely studied class of modified gravity theories involves replacing the scalar curvature of the Einstein-Hilbert action by a more general function $f(R)$.
This leads to the equations of motion with fourth order derivatives in contrast to the second order differential equations of GR.
A large number of different $f(R)$ theories have been under scrutiny (see {\it e.g.} \cite{flanagan2003, Hu2007, Starobinsky2007, Tsujikawa2007, Appleby:2007vb}).

The most important characteristic of these models is the generated accelerated expansion. The first proposed models were quickly discarded as problems arose with stability and solar system constraints. As studies have found theoretical and observational constraints on possible models ({\it e.g.} \cite{Dolgov2003}), the viable models have become more refined. 

The challenge of $f(R)$ theories is surviving the strict solar-system bounds and simultaneously creating the accelerated expansion at late times. These theories can be interpreted as introducing a scalar degree of freedom \cite{Chiba2003} which may cause considerable deviation from GR around the Sun. A viable model should therefore include a mechanism to hide the new effects on high curvature regimes \cite{Khoury:2003rn}. This is achieved by $f(R)$ models where the squared mass of the scalar is large in the large curvature region \cite{Song:2006ej}. The same condition is set by the high-redshift observations of the cosmic microwave background (CMB) \cite{Amendola:2006kh}.

Besides explaining the accelerated expansion, $f(R)$ theories have been shown to have other benefits. It may be related to the early inflationary expansion of the Universe \cite{PhysRevD.77.026007}. Moreover, it has been shown that with modified gravity the rotation curves of spiral galaxies and the halos of the core clusters could be explained without dark matter \cite{Boehmer:2007kx,Capozziello:2006ph,Capozziello:2008ny}, and $f(R)$ theories have also been shown to give possible solutions to problems related to other objects such as neutron stars \cite{Cooney:2009rr}.

In the present article, we study some of the most successful $f(R)$ theories by considering the structure formation. This is done using the Jeans instability analysis of self-gravitating systems, where {\it e.g.} star formation can be examined. Instabilities in self-gravitating systems were first studied by Jeans \cite{jeans_original}. As this was before the advent of
the general relativity, the analysis was restricted to non-relativistic, Newtonian gravity. Later on, Jeans analysis has been upgraded to use GR and some works have further extended it to modified gravity \cite{Capozziello:2011gm,Roshan:2014mqa}.

For $f(R)$ models, it is possible to find further constraints for viable models \cite{Capozziello:2011gm}. We generalize the method and apply it to molecular clouds. These could offer a new class of objects to measure the viability of $f(R)$ models as there is ample observational data on large molecular clouds \cite{RomanDuval:2010nm}. However, the masses of large clouds are several magnitudes higher than the Jeans masses, but in the smaller Bok globules the cloud masses are close to the well-known Jeans limit. Therefore, the $f(R)$ modified Jeans mass and the standard Jeans mass may yield different predictions on whether a globule is about to collapse.

\section{Equations of motion}
In $f(R)$ gravity the Einstein-Hilbert Lagrangian is not set {\it a priori} to be the linear $f(R)=R$. The function $f(R)$ is an analytic function of the curvature scalar $R$. If we set the requirement of no higher derivatives than second degree, the function reduces to $R$ and the Einstein-Hilbert Lagrangian is obtained. With a generalized function it is possible to find a better match to observations than the simplest choice, $f(R)=R$.

We consider general $f(R)$ modifications to the Einstein-Hilbert action,
\be
\mathcal A =\frac 1{2\chi}\int d^4x\sqrt{-g}\Big(f(R)+2\chi \mathcal L_m\Big),
\ee
where $\chi=\frac{8\pi G}{c^4}$ is the coupling of gravitational equations. The latter term $\mathcal L_m$ is the minimally coupled matter Lagrangian. With $f(R)=R$ the action would reduce to the standard Einstein-Hilbert action. There are several restrictions to the possible form of the function $f(R)$ which are further discussed in section \ref{constraints}. The signature of the metric is $-,+,+,+$, the Riemann curvature tensor is $R^\alpha_{\beta\mu\nu}=\partial_\mu\Gamma^{\alpha}_{\beta\nu}-\partial_\nu\Gamma^{\alpha}_{\beta\mu}+\Gamma^\alpha_{\kappa\mu}\Gamma^{\kappa}_{\beta\nu}-\Gamma^\alpha_{\kappa\nu}\Gamma^\kappa_{\beta\mu}$ and the Ricci tensor is $R_{\mu\nu}=R^\alpha_{\mu\alpha\nu}$. Using standard metric variational techniques we find the field equations and the trace equation
\ba\label{eqmo}
f'(R)R_{\mu\nu}-\frac 12f(R)g_{\mu\nu}-\nabla_\mu\nabla_\nu f'(R)+g_{\mu\nu}\square f'(R)&=\chi T_{\mu\nu} \\
3\square f'(R) + f'(R)R-2f(R)&=\chi T,
\end{align}
where $T_{\mu\nu}=-\frac 2{\sqrt{-g}}\frac{\delta\sqrt{-g}\mathcal L_m}{\delta g^{\mu\nu}}$ is the energy-momentum tensor and $T=T^\alpha_\alpha$. A prime is used to denote the derivatives with respect to $R$. As we are about to examine collapsing molecular clouds ({\it i.e.} relatively thin matter), we are going to use a weak field approximation for the metric. The background is assumed to be of de Sitter form with perturbations added to the diagonal elements \cite{mukhanov1990}. In polar coordinates $x^\mu=(t,r,\theta,\phi)=(t,\mathbf x)$ we have a diagonal metric up to $\mathcal O(3)$ with
\ba\label{metric}
g_{00}&=-\big(h(r)+2\phi(t,\mathbf x)\big), \\
g_{11}&=1/h(r)+2\Psi(t,\mathbf x), \\
g_{22}&=(1+2\Psi(t,\mathbf x))r^2, \\
g_{33}&=(1+2\Psi(t,\mathbf x))r^2\sin\theta
\end{align}

The expansion parameter is $v$, the velocity of a test particle\footnote{The expansion parameter can be equivalently $c^{-1}$ as the velocity appears as a combination $v/c$. As it is customary to set $c=1$, we prefer $v$ as the expansion parameter.} \cite{weinberg1972}. Note, that this form of metric tensor corresponds to the first order post-Newtonian approximation in quasi-Minkowskian coordinates.

The perturbation in the temporal component of the metric, $\phi(t,\mathbf x)$ corresponds to the Newton gravitational potential. It can be further broken up as $\phi(t,\mathbf x)=\phi_0+\Phi(t,\mathbf x)$, where the constant $\phi_0$ refers to the local environment around the object in question. For example, for the galactic potential we would have $\phi_0\approx 2\times 10^{-6}$ \cite{Jain:2012tn} (with $c$ set to unity). For large scale considerations this constant term must be discarded as there is no constant background. Due to the derivatives involved in calculating the curvature tensor and scalar, this constant term does not affect curvature.

The non-diagonal elements must have odd powers as the time reversal transformation (as well as other coordinate reflections) should change the sign. Therefore, they are at least of order $\mathcal O(3)$. For the case of a weak field limit, terms of $\mathcal O(3)$ and higher are discarded. The Ricci scalar can be expanded around the background as
\be
R\simeq R_0+R^{(2)}(t,\mathbf x)+\mathcal O(4),
\ee
where $R^{(n)}$ denotes that the quantity is $\mathcal O(n)$. For de Sitter background $R_0=4\Lambda$. As the derivatives of $f(R)$ appear in the equations of motion, we need an expansion for this function as well
\be
f^n(R)\simeq f^n\big(R_0+R^{(2)}+\mathcal O(4)\big)\simeq f^n(R_0)+f^{n+1}(R_0)R^{(2)}+\mathcal O(4),
\end{equation}  
which can be iterated for the desired order. In our case the first order is sufficient {\it i.e.} $f'(R)\simeq f'(R_0)+f''(R_0)R^{(2)}+\mathcal O(4)$. Not all the characteristics of $f(R)$ models manifest at this order. However, in the scope of this paper, we concentrate on the lowest order effects on stability. If differences between GR and $f(R)$ appear in a lower order, they are not likely to be canceled in higher orders. Inserting these into \eqref{eqmo} we have in second order
\ba
f'(R_0)\Big(R^{(2)}_{tt}+\frac{R^{(2)}}2\Big)-f''(R_0)\nabla^2R^{(2)}+f'(R_0)R_{tt}^{(0)}-\frac{f(R_0)}2 g_{tt}=\chi T^{(0)}_{tt}, \label{eqmo1}\\
3f''(R_0)\Big(\nabla^2-\partial_0^2\Big)R^{(2)}-f'(R_0)R^{(2)}+f'(R_0)R_0-2f(R_0)=\chi T^{(0)} \label{eqmo2},
\end{align}
where $\nabla^2$ is the spatial flat Laplacian. The flatness of the Laplacian is due to the quasi-Minkowskian nature of the metric \cite{weinberg1972}. In contrast to \cite{Capozziello:2011gm}, we have included the time derivatives as collapsing clouds are time-dependent and dynamic. For a viable $f(R)$ theory to have a de Sitter solution, $f'(R)R=2f(R)$ must hold. This is in order to achieve the cosmic acceleration. Observations ({\it e.g.} Planck results \cite{Ade:2015xua}) show that the current evolution of the Universe is close to de Sitter behaviour. Therefore, the $\Lambda$ background and the solution associated to it must exist as well as the solutions for the matter on the foreground. For the background the solution is $f'(R_0)R=2f(R_0)$. The deviation of the current curvature from the de Sitter space is due to the matter content, which is caused locally, as the curvature scalar is a local quantity. In the first order expansion \cite{Capozziello:2011wg} the Birkhoff theorem is valid and allows us to separate the background. This leads to cancellation of the last two terms in \eqref{eqmo2}. 

With the same substitutions and $R_{tt}^{(0)}=-\Lambda$ we look at the last two terms in \eqref{eqmo1} to find them equal to $2f'(R_0)\Lambda(\phi_0+\Phi(t,\mathbf x))$. The dynamic term would be of higher order. The constant term is clearly small as well but deserves a closer look. As part of the background it effectively works as a source of curvature. In that sense it is best compared to the other sources, {\it i.e.} the energy momentum tensor on the right side of the equation.

The perturbation terms of the Ricci scalar and the $tt$ component of the Ricci tensor can be calculated from the perturbed metric \eqref{metric}
\ba
R^{(2)}&=6\ddot \Psi-2\nabla^2\Phi-4\nabla^2\Psi, \label{ricci1} \\ 
R_{tt}^{(2)}&=\nabla^2\Phi-3\ddot \Psi \label{ricci2}.
\end{align}
For the equations of motion the energy-momentum tensor must be defined. We use the perfect fluid form
\be
T_{\mu\nu}=(\rho+p)u_\mu u_\nu -pg_{\mu\nu}.
\ee
with $p$ being the pressure and $\rho$ the mass density. As the molecular clouds consist of dust, we further set $(p=0)$ and obtain
\ba\label{eqmop}
-2f'(R_0)\nabla ^2 \Psi +f''(R_0)\big(2\nabla^4\Phi+4\nabla^4\Psi-6\nabla^2\ddot\Psi\big) =\chi \rho +2f'(R_0)\Lambda\phi_0,&\\
f''(R_0)\big(5\nabla^2\ddot\Psi+\nabla^2\ddot\Phi-\nabla^4\Phi-2\nabla^4\Psi\big)-f'(R_0)\big(\ddot\Psi+\frac{\nabla^2\Phi}{3}+\frac{2\nabla^2\Psi}{3}\big) =-\frac{\chi \rho}{6}&.
\end{align}
On the right side of the equations, the elements of the energy-momentum tensor are of the zeroth order due to the coupling constant $\chi$ being second order. We have omitted the term $-18f''(R_0)\ddddot\Psi$ as it is of higher order due to the multiple time derivatives and, therefore, being smaller. 

The relation of the time derivatives and spatial derivatives and the order merits a mention. The derivatives have an effect on the expansion order (see \cite{weinberg1972}, chapter 9). The time derivatives raise the the order since
\ba
\frac\partial{\partial_i}\sim \frac 1r, \\
\frac\partial{\partial_t}\sim \frac vr.
\end{align}

This calls for a question whether also second order perturbations should be included. In the linear Jeans analysis the perturbations examined are arbitrarily small and we may assume $\Phi\gg\Phi^2$. It is also worth mentioning that while first order perturbations can cause second order perturbations to appear, the second order perturbations cannot cause first order perturbations to appear. For this reason we have not included quadratic terms in the Ricci tensor \eqref{ricci1} and scalar \eqref{ricci2}.

We can rescale \eqref{eqmop} by dividing it with $f'(R_0)$ on both sides. This way, the effect of a chosen $f(R)$ model is incorporated into the ratio $f''(R_0)/f'(R_0)$ and leads to a scaled gravitational constant $\chi/f'(R_0)$. For simplicity we aim to set $f'(R_0)=1$. 

For GR, we have $f'(R)=1$ and measurements of the gravitational constant indicate its relative error is $\delta G/G<1.2\times 10^{-4}$ \cite{Agashe:2014kda}. Using the expansion $f'(R)\simeq f'(R_0)+f''(R_0)R^{(2)}$ the solar system results and the $f'(R_0)$ can be related. Since the perturbation $R^{(2)}\ll 1$ by definition, the second term would be small as well, unless $f''(R_0)\gg 1$, which would cause physical anomalies. Therefore, the effective deviation
\be
\frac{\delta G}G=\Big|\frac{f'(R)-1}{f'(R)}\Big|<1.2\times 10^{-4},
\end{equation}
and we set $f'(R_0)=1$.

The value of the cosmological constant should be extremely small \cite{Weinberg:1988cp}. Therefore, the second term on the right side of \eqref{eqmop} would be small as well. For dust clouds, such as the Bok globules, the ratio of the terms is $\Big|\frac{\Lambda\phi_0}{\chi\rho}\Big|\sim 10^{-5}$. As the the galactic potential originates from the matter content of the Milky Way, the second term corresponds to a constant part of $\rho$. In the scale of a single dust cloud the background $\rho$ is constant. It can also be argued that to this order the Birkhoff theorem can be applied in this weak field approach. In low orders of expansion and a static Ricci scalar (here, the galactic background), the Birkhoff theorem is valid \cite{Capozziello:2007ms,Capozziello:2011wg}. Therefore, the net effect on the globule would be negligible. On these grounds we remove the second term in the following treatment.

For $f''(R_0)=0$ with a static potential $\ddot\Phi=0$ the standard Poisson equation of $\nabla^2\Phi=4\pi G\rho$ is reached. In the Newtonian case the only perturbation considered is static $\Phi$, omitting the spatial perturbation $\Psi$.

\section{Collapse in a self-gravitating collisionless system}

A self-gravitating system of particles in equilibrium is described by a time-independent distribution function $f_0(\mathbf x,\mathbf v)$ and a potential $\Phi_0(\mathbf x)$. They are the solutions of the collisionless Boltzmann equation and the Poisson equation
\ba
\nabla^2\Phi(\mathbf x,t)=4\pi G\int f(\mathbf x,\mathbf v, t)d\mathbf v, \\
\frac{\partial f(\mathbf x,\mathbf v, t)}{\partial t}+(\mathbf v\cdot\nabla_x)f(\mathbf x,\mathbf v,t)-(\nabla\Phi\cdot\nabla_v)f(\mathbf x,\mathbf v,t)=0. \label{bol}
\end{align}
Here $\mathbf v$ and $\mathbf x$ are spatial velocity and position vectors and the $\nabla$ operates in the three spatial dimensions. In the Newtonian limit $\Phi_0$ is just the gravitational potential of the metric \eqref{metric}.

Following standard methods ({\it e.g.} \cite{BT2}) we linearize these two equations and write them in Fourier space to obtain (for clarity we omit writing the variables)
\ba
-i\omega f_1+\mathbf v\cdot(i\mathbf k f_1)-(i\mathbf k\Phi_1)\cdot\frac{\partial f_0}{\partial \mathbf v}=0, \label{bolf} \\
-k^2\Phi_1=4\pi G\int f_1d\mathbf v.
\end{align}
We can now solve for
\be
f_1=\frac{\mathbf k\cdot \frac{\partial f_0}{\partial\mathbf v}}{\mathbf v\cdot \mathbf k-\omega}\Phi_1 \label{bolff1}.
\ee

For the purposes of Jeans analysis, we need to consider small perturbations to the equilibrium and linearize the equations of motion. We write the mass distribution function $\rho=\int f(\mathbf x, \mathbf v, t)d\mathbf v$ in \eqref{eqmop} and write the equations in Fourier space to get the following equations
\ba
k^2\Psi_1+k^2\alpha (k^2\Phi_1+2k^2\Psi_1-3\omega^2\Psi_1)&=4\pi G\int f_1 \ d\mathbf v, \label{fourier1}\\
3k^4\alpha \big(\frac{\omega^2}{k^2}(5\Psi_1+\Phi_1)-\Phi_1-2\Psi_1\big)-k^2\Phi_1-2k^2\Psi_1+3\omega^2\Psi_1&=-4\pi G\int f_1 d\mathbf v .
\end{align}
We have denoted $f''(R_0)=\alpha$, which conveys the effects of $f(R)$ theories. From these two equations we can solve for
\be
\Psi_1=-\frac{k^2(1+2\alpha k^2-3\alpha\omega^2)}{(1+4\alpha k^2)(k^2-3\omega^2)}\Phi_1, \label{psi1}
\ee
which can be inserted back into \eqref{fourier1} to obtain
\be
\frac{k^4\Big(1+3\alpha(k^2-\omega^2+3\alpha\omega^4)\Big)}{(1+4\alpha k^2)(k^2-3\omega^2)}\Phi_1=-4\pi G\int f_1d\mathbf v.
\ee
With the solved linearized matter distribution \eqref{bolff1} we reach the dispersion relation
\be
4 \pi G\int\Big(\frac{\mathbf k\cdot \frac{\partial f_0}{\partial\mathbf v}}{\mathbf v\cdot \mathbf k-\omega}\Big)d\mathbf v+\frac{k^4\Big(1+3\alpha(k^2-\omega^2+3\alpha\omega^4)\Big)}{(1+4\alpha k^2)(k^2-3\omega^2)}=0.\label{dispers}
\ee
In the standard case, $\alpha=0$, the limit for instability is found at 
\be
k^2_{\omega=0}=\frac{4\pi G\rho_0}{\sigma^2}\equiv k^2_J. \label{mjdef}
\ee
Which is called the Jeans wavenumber. With this we can define the Jeans mass as the mass which was initially inside a sphere of diameter $\lambda_J$:
\be
\lambda_J^2\equiv\frac{4\pi^2}{k_J^2}=\frac{\pi\sigma^2}{G\rho_0}, \ \ \
M_J\equiv\frac{4\pi\rho_0}3\Big(\frac{\lambda_J}2\Big)^3=\frac\pi 6\sqrt{\frac 1{\rho_0}\Big(\frac{\pi\sigma^2}G\Big)^3}.\label{mj}
\ee
The Jeans length $\lambda_J$ is the limit beyond which the perturbations are unstable, experiencing exponential growth. On the other hand $\lambda_J<2\pi/k_J$ perturbations in stellar systems the response is strongly damped even though the system contains no friction \cite{BT2}. The Jeans mass, however, is more useful for our purposes of probing the stability of interstellar clouds. If the mass of the cloud exceeds Jeans mass $M_J$, it will collapse.

\section{Jeans instability limit in the $f(R)$ case}\label{frlimit}
To discuss the case $f''(R_0)\neq 0$ we return to the dispersion relation \eqref{dispers}. It can be recast into (see the appendix \ref{append} for details)
\be
1-\sqrt\pi xe^{x^2}\text{erfc}(x)=\frac{k^4\Big(1+3\alpha(k^2+\omega_I^2+3\alpha\omega_I^4)\Big)}{k^2_J(1+4\alpha k^2)(k^2+3\omega_I^2)}, \label{integrated}
\ee
with $x=\frac{|\omega_I|}{\sqrt 2 k\sigma}$ and $\omega_I=-i\omega$. The left side is a bounded monotonously decreasing function in respect with $x$. The limit for instability is found at $\omega=\omega_I=x=0$ where the left side of \eqref{integrated} reduces to unity.
\be
k_J^2-\frac{k^2(1+3\alpha k^2)}{1+4\alpha k^2}=0. \label{kjkfr}
\ee
which can be simplified into $\alpha k^4+ \big(1-4\alpha k_J^2\big)k^2-k_J^2=0$. If $\alpha=0$, the equation would be of lower order and produce only the standard solution. If $\alpha\neq 0$, several solutions are possible. Besides the apparently excluded case $-1/3k^2<\alpha<-1/4k^2$ where $k^2$ would be negative, \eqref{kjkfr} can be solved for
\be
k^2=k^2_\pm=\frac{-1+4\alpha k_J^2\pm\sqrt{1+4\alpha k_J^2+16\alpha^2 k_J^4}}{6\alpha}. \label{kwave}
\ee
With \eqref{mj} we can write the $f(R)$ modified Jeans mass as
\be
\tilde M_{J\pm}=\Big(\frac{6\alpha k^2_J}{-1+4\alpha k^2_J\pm\sqrt{1+4\alpha k^2_J+16(\alpha k^2_J)^2}}\Big)^{3/2}M_J\equiv\beta_\pm M_J \label{frjeans}.
\ee
To reach a real mass the expression inside the brackets must be positive. It is apparent that for the $\beta_-$ solution, we must have $\beta<0$ to avoid complex masses. This equation describes the relation of standard GR Jeans mass for a self-gravitating stellar system and one described with $f(R)$ gravity.

\subsection{Solutions for non-zero $\omega$}
The dispersion equation \eqref{integrated} can be solved with $\omega=0$ to get the instability limit \eqref{kwave} but there might be other solutions as well. The physical meaning of these $\omega\neq 0$ solutions merits a brief inspection.

Solving the dispersion equation for $k(\omega)$ is difficult and unnecessary. Examination of the derivatives on both sides of the equation \eqref{integrated} is sufficient to reveal the existence of at least one non-zero solution.

These non-zero $\omega$ solutions are also present in the standard case of $\alpha=0$. If we examine a case in which $k=k_J-\delta k$, we notice that this corresponds to a mass slightly over the Jeans mass, $M=M_J+\delta M$. In the mean time, this $k$ would require a non-zero $\omega$ for the dispersion relation to hold. The physical interpretation is, that when the object ({\it e.g.} a dust cloud) has a mass exceeding the Jeans mass, it will collapse even if it has a small initial radial velocity.  

These non-zero solutions would appear at high values of $\omega$. However, the original expansion around the background would break upon leaving the neighborhood of $\omega=0$. For this reason it is possible to examine only the case of $\omega=0$.

\subsection{Characterization of $k_-$ and $k_+$}
In the standard case of Jeans stability analysis, there is only one limit for unstable modes. With a more general $f(R)$ case the situation changes and there are possibly two limits for instability \eqref{frjeans}. The appearance of other limits is in a way expected as $f(R)$ theories allow for an additional degree of freedom (see {it e.g.} \cite{faraoni04, Sotiriou:2008rp}) which is perhaps best illustrated through the scalar-tensor theory equivalence. The physical significance of these two limits must be addressed.

The standard Jeans mass should be recovered with $f(R)=R$ which corresponds to $\alpha=0$. Upon examining \eqref{kwave}, this is leads to $k_+^2=k_J^2$ and $k_-^2$ has the asymptotic behaviour $-1/3\alpha$, which diverges. Therefore, GR would have the standard Jeans mass and there would be no other meaningful limit, as is to be expected. With this observation the solution $k_+$ can be labeled as the generalization of the standard Jeans wave number. 

The addition of a more complicated $f(R)$ leads to different results depending on the sign of $\alpha$. For $\alpha>0$, the $k_-^2$ solution would translate to a negative Jeans mass. The result is one modified Jeans mass. This modified mass is lower than the standard one. For a dust cloud this means assisted star formation.

For negative $\alpha$ the situation is more interesting as there are two positive solutions for $k^2$. The new solution $k_-$ would produce a considerably lower limit, converging to zero at $\alpha=0$. The physical meaning of this limit must be addressed. If it translates into a lower limit for collapse, the effects for {\it e.g.} star formation would be observable. The $k_+$ solution refers to higher Jeans mass than standard case. In this case the expansion of the cosmic background counteracts the collapse.

The limit for instability was found earlier by setting $\omega=0$, {\it i.e.} the mass distribution is time-independent, stable. All the contracting modes must fulfill the dispersion equation \eqref{integrated}. This is the case for $k<k_+$ as in the standard case. However, for $k>k_-$ there are no solutions \eqref{integrated}. The left side of the equation is monotonous and has an upper limit 1. The right side has values over 1 when $\alpha<0$ and $k>k_-$.

As the dispersion relation does not hold with $k>k_-$, the limit $k_-$ is not a limit for collapsing modes. However, the with $\omega=0$ this corresponds to a stable mode. There are three forces in action in a collapse scenario, mass causing the collapse, thermal movement keeping it apart and, with an $f(R)$ model, the background effect. With $k>k_+$ the mass is not sufficient to counteract the thermal movement, so the the mass distribution, {\it e.g.} a dust cloud, starts to oscillate, with non-zero $\omega$.

As can be observed from \eqref{integrated}, temperature does not appear explicitly in the dispersion relation with $\omega=0$. Thus, for $k_-$ the balancing forces are mass and the expansion of the background. With non-zero $\omega$ the equation will not hold. It would require extreme fine-tuning to reach this state and would be lost due any external perturbation. Therefore, it will not be physically meaningful. One more reason to discard the $k_-$ solution is the Dolgov-Kawasaki instability, which is covered in section \ref{constraints}. With the Dolgov-Kawasaki instability and non-negativity of the Jeans mass both $\alpha>0$ and $\alpha<0$ are denied for $k_-$. For these reasons we restrict to $\beta_+$ and $\tilde M_{J+}$ for the following treatment and omit the subscript signs.

Even though in our case there remains only the modified Jeans mass and one instability, in other situations these new instabilities might endure. In \cite{Arbuzova:2015uga} instabilities and collapse were studied in oscillating backgrounds. These situations demand the inclusion of higher order derivative terms in the equations of motion. This difference allows for different instabilities to manifest in situations like black hole formation.

\section{Comparison of Jeans masses in GR and $f(R)$ models}
We derive the range within which $f(R)$ models fall compared to the GR. Using the definition of Jeans mass \eqref{mj} and the derived $f(R)$ Jeans mass \eqref{frjeans} we can write
\be
\tilde M_J=\beta\frac\pi 6\sqrt{\frac 1{\rho_0}\Big(\frac{\pi\sigma^2}G\Big)^3}.
\ee
If we are to examine star formation, $\rho_0$ is the interstellar medium density (ISM) and $\sigma$ is the velocity dispersion of particles due to temperature,
\be
\rho_0=m_Hn_H\mu, \ \sigma^2=\frac{k_BT}{m_H}
\ee
with $n_H$ being the number of particles, $\mu$ the mean molecular weight (check \cite{przybilla} for values in molecular clouds), $k_B$ the Boltzmann constant and $m_H$ the proton mass. With these we compute the behaviour of $\tilde M_J$ for a given $f(R)$ model described by $\beta$
\ba
\tilde M_J&=\beta\frac{\pi T^{3/2}} {6m_H^2}\sqrt{\frac 1{n_H\mu}\Big(\frac{\pi k_B}G\Big)^3}.
\end{align}
The asymptotic value for $\beta\to\infty$ and $\beta\to 0$ are easily found to be
\begin{align}
\lim_{\beta\to\infty}\tilde M_J & = (3/4)^{3/2}M_J \\
\lim_{\beta\to 0}\tilde M_J&=M_J \\
\lim_{\beta\to-\infty}\tilde M_J & = \infty
\end{align}
Therefore for $\beta$, inserting the values of the constants,
\be
\frac{\tilde M_J}{M_J}\in(0.649519,1].
\ee
We see now, that $f(R)$ gravity can cause a considerably lower limit for gravitational collapse. For theories with positive $\beta_+$, this would assist in star formation. For negative values, the effect is inverse and would lead to reduced star formation. In the following section we will use compare these limits to observations of Bok globules.

\section{Jeans mass limit in Bok globules}
Bok globules are clouds of interstellar gas and dust. These dark clouds are relatively hard to spot and therefore all the observed globules are located nearby, on the galactic scale. The cloud cores are cold at temperatures of around $10$K. Most of the observed globules are isolated and of simple shape. Masses of these globules tend to be less than $100M_\odot$ with many around $10M_\odot$. This is considerably less than the large molecular clouds in the Milky Way, which are several orders of magnitude greater (e.g. \cite{RomanDuval:2010nm}).

Bok globules have masses and the corresponding Jeans masses of the same order. Therefore, we can observe Bok globules for which the classic Jeans mass and the $f(R)$ corrected Jeans mass give a different prediction for stability. There are observations of hundreds of globules \cite{1995MNRAS.276.1052B, 1995MNRAS.276.1067B, CB}, with estimates of the total number of globules in the Milky Way at tens of thousands \cite{1980fleck}. It has been found \cite{1990ApJ...365L..73Y, 1995MNRAS.276.1052B, schmalzl2014} that most of the Bok globules experience star formation with one or more star forming cores. 


The formation of the globules themselves is a process not well understood. It is possible that they form as condensations of diffuse gas in relative isolation. Another explanation is that the globules form as dense cores of larger interstellar clouds \cite{1999ApJ...524..923N}. This agrees with the greater density. The presence of large external masses of stellar winds may also play a role in starting the collapse and star formation. 



The observation of Bok globules is somewhat problematic. Extensive tables on their properties are not yet readily available. The kinetic temperatures are calculated from ammonia observations \cite{1995MNRAS.276.1067B}. With excitation and kinetic temperatures known, the molecular hydrogen number density can be found. The reported masses are calculated for the globule cores assuming homogeneous distribution and spherical symmetry. However, the majority of globules as a whole are not spherical but elliptical \cite{1995MNRAS.276.1052B}.

For these reason the physics of Bok globules are not yet completely understood. For our purposes of looking into the stability of the clouds, the observations are sufficient as a demonstration for the viability of the method. Better accuracy in measurements of density and temperature would provide for a more accurate study. A more accurate modeling of the collapse would also take into account other forces, such as magnetic or turbulent. Nevertheless, the following will serve as a feasibility study on using Bok globules for constraints.

The chosen globules are the ones in \cite{1995MNRAS.276.1067B} which have calculated kinetic temperatures, hydrogen number densities and masses. The dark cloud names are those given in \cite{1986hartley}. For our purposes of comparing the Jeans masses, we use the observational data from several Bok globules in \cite{2005Kandori}.
\begin{table}
\scriptsize
\label{table1}
\begin{tabular}{|c|l|l|l|l|l|c|c|c|} \hline
 &$T\text{[K]}$ & $n_{H_2}\text{[cm$^{-3}$]}$ & $M$  & $M_J$ & $\tilde M_J$ & Prediction & Stability \\ \hline
CB 87	& 11.4 &	$(1.7\pm 0.2)\times 10^4$	& $2.73\pm 0.24$ & 9.6 & 6.2 & stable & stable \\ \hline
CB 110 & 21.8 & $(1.5\pm 0.6)\times 10^5$ & $7.21\pm 1.64$ & 8.5 & 5.5 & MD	& unstable \\ \hline
CB 131 & 25.1 & $(2.5\pm 1.3)\times 10^5$ & $7.83\pm 2.35$ & 8.1 & 5.3 & MD & unstable \\ \hline
CB 134 & 13.2 & $(7.5\pm 3.3)\times 10^5$ & $1.91\pm 0.52$ & 1.8 & 1.2 & unstable & unstable \\ \hline
CB 161 & 12.5 & $(7.0\pm 1.6)\times 10^4$ & $2.79\pm 0.72$ & 5.4 & 3.5 & stable & unstable \\ \hline
CB 184 & 15.5 & $(3.0\pm 0.4)\times 10^4$ & $4.70\pm 1.76$ & 11.4 & 7.4 & stable & unstable \\ \hline
CB 188 & 19.0 & $(1.2\pm 0.2)\times 10^5$ & $7.19\pm 2.28$ & 7.7 & 5.0 & MD & unstable \\ \hline
FeSt 1-457 & 10.9 & $(6.5\pm 1.7)\times 10^5$ & $1.12\pm 0.23$ & 1.4 & 0.94 & MD & unstable\\ \hline
Lynds 495 & 12.6 & $(4.8\pm 1.4)\times 10^4$ & $2.95\pm 0.77$ & 6.6 & 4.3 & stable & unstable \\ \hline
Lynds 498 & 11.0 & $(4.3\pm 0.5)\times 10^4$ & $1.42\pm 0.16$ & 5.7 & 3.7 & stable & stable \\ \hline
Coalsack & 15 & $(5.4\pm 1.4)\times 10^4$ & $4.50$ & 8.1 & 5.3 & stable & stable \\ \hline
\end{tabular}
\caption{For selected Bok globules we present the name, kinetic temperature, particle number, mass, conventional Jeans Mass, lowest possible Jeans mass due to $f(R)$ gravity, stability prediction from Jeans mass and stability reported in \cite{2005Kandori}. Masses are in the units of solar masses $M_\odot$. MD stands for the case where the prediction depends on the chosen $f(R)$ parameter.}
\end{table}

The listed errors in parameters are 1$\sigma$ deviations. The reported temperatures are effective temperatures which include the kinetic and the turbulent nonthermal component \cite{2005Kandori}.

We notice that the modified $f(R)$ Jeans masses offer better agreement than the conventional Jeans masses. In four of the globules, (CB 110, CB 131, CB 188 and FeSt 1-457) the mass exceeds the modified Jeans mass for some of the theories but not the conventional Jeans mass. In fact only one of the observed globules, CB 134, has a mass exceeding the conventional Jeans mass. Clearly, having a lower limit for the collapse due to $f(R)$ gravity agrees with observations.

There is a disagreement on three globules, (CB 161, CB 184 and Lynds 495). Even though the mass of the globule is lower than the critical mass, a collapse can occur. In these cases however, it is due to some external force, {\it e.g.} a shock wave from a supernova.

According to \cite{2005Kandori} the globules with disagreement (CB 161, CB 184 and Lynds 495) in the prediction are "marginally unstable" which is a state with a considerably longer lifetime. These perturbations take considerably more time to dissipate. During that time some external force to begin the collapse is more likely to take place. Therefore, it is not directly contradictory to our findings. 

\begin{figure}\label{plotpic}
\includegraphics[scale=0.8]{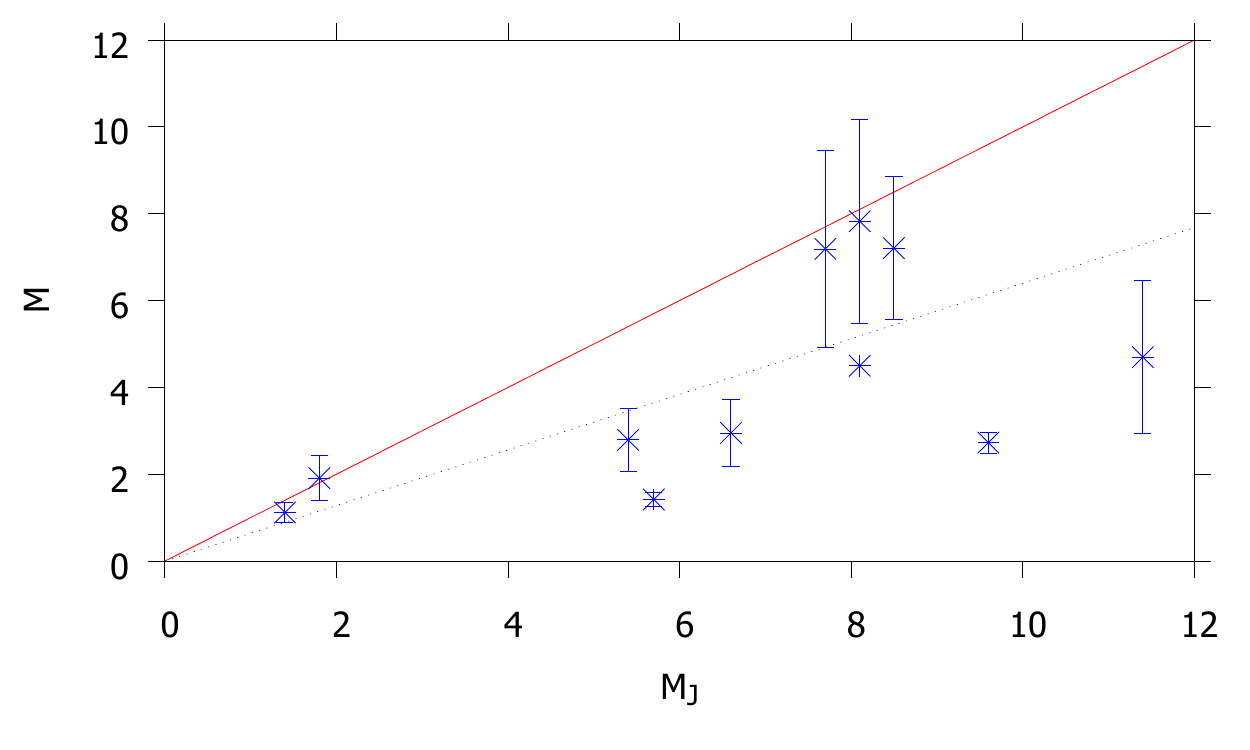}\label{fig-n10rs}
\caption{The masses of the examined Bok globules are presented with the graphs of the conventional Jeans mass, solid line, and the lowest $f(R)$ modified dotted line. The units on both axes the solar mass units. The part above the solid line is the stable zone, whereas the one below the dotted line is the collapsing region. The part between the lines depends on the chosen $f(R)$ model.}
\end{figure}

The globule CB 188 is found to have a protostar. As the prediction for this globule depends on the chosen $f(R)$ theory, it supports the $f(R)$ modified Jeans mass which can also been seen in \ref{plotpic}. The conventional zone for collapse is within the error of the measured mass of CB 188, but the collapse better fits the modified Jeans mass. For the globule mass to be sufficient the Jeans mass modification coefficient should be $\beta < 0.93$. With the definition of the Jeans mass \eqref{mjdef} we have for this globule
\be
f''(R_0)> 0.052k_J^2=0.052\frac{4\pi G m_H^2 n_H\mu}{k_B T}\approx 1.3\times 10^{-31}\text{m}^{-2}.
\ee

The most interesting globule in the sample is FeSt 1-457, for which the non-altered Jeans mass  is well above the observed mass even with the error taken into account. For FeSt 1-457 the coefficient would be the even lower $\beta < 0.78$ at the best estimate and the same as for CB 188, $\beta < 0.93$, for the upper limit. With the upper limit we would have a constraint for $f(R)$ models
\be
f''(R_0)>1.3\times 10^{-31}\text{m}^{-2}.\label{cond}
\ee
	
In this sense the $f(R)$ modified Jeans mass appears to better predict the collapse of globules. As the mass necessary to make clouds collapse is less, this has an effect on the forces holding the collapse at bay. This would imply that the counteracting forces, {\it e.g.} turbulence, do not need to be as strong in molecular clouds as with conventional Jeans mass. 

We stress that this is not a stringent constraint for $f(R)$ theories. As it is produced by a single data point, it rather serves as a feasibility study. With a larger dataset, preferably with smaller error bars, it is possible to find a constraint for $f(R)$ models and other types of modified gravity as well.

\section{Constraints for $f(R)$ models}\label{constraints}
In this section we take a look at specific $f(R)$ models which are considered viable. The treatment of the previous chapter is subjected to these models. In the literature, there are numerous general constraints on viable models. First we take a look at these known constraints. 

There are several necessary conditions for $f(R)$ models to satisfy \cite{DeFelice:2010aj}. Some of the conditions are based on mathematical properties and some are related to observations of the Universe. Some of the most simple and fundamental ones are the ghost and the Dolgov-Kawasaki criteria. To avoid ghosts and anti-gravity the condition $f'(R)>0$ is necessary \cite{Starobinsky2007}. The Dolgov-Kawasaki singularity is avoided with $f''(R)>0$ \cite{Dolgov2003}. 

The Dolgov-Kawasaki criterion effectively rules out the somewhat unambiguous solutions of $k_-$. It should be noted that this criterion must be satisfied at $R\geq R_a$, where $R_a$ is the present day curvature. On examining perturbations on a Minkowskian background as in \cite{Capozziello:2011gm}, these constraints do not need to hold and the $k_-$ solution is not ruled out.



Most cosmological constraints come from far-away objects such as supernovae \cite{riess98} and large scale structures \cite{Perlmutter1999, Bunn1996} but nearby objects can be considered as well. Several types of astrophysical objects have been considered for constraints in the literature \cite{Jain:2012tn, Schmidt:2008tn}. These include cepheids, red giant stars, water masers and relatively closer dwarf galaxies \cite{Vikram:2013uba, Jain:2011ji,Hui:2009kc}.

Next we will examine some specific $f(R)$ models and the constraints set by the globule observations.

\subsection{Hu-Sawicki model}
The Hu-Sawicki models produce the accelerated expansion and satisfy both cosmological and solar-system constraints. There are three parameters, for which there are some constraints. The $f(R)$ function reads as 
\begin{equation}
f(R)=R-\lambda R_c\frac{(R/R_c)^{2n}}{(R/R_c)^{2n}+1}, \ n,\lambda, R_c>0.
\end{equation}
The critical curvature $R_c$ is of the order of the present day curvature. The larger the $n$ the longer the model mimics $\Lambda$CDM. It has been found that there is also a lower limit for $n>0.905$ \cite{DeFelice:2010aj} and in \cite{Okada:2012mn} it is found that for $n=1$ the $\lambda$ must be large $\lambda\gg 20$.

The only non-zero values for $f''(R_0)$ are with $n=1/2$ and $n=1/4$. These are both ruled out by the condition $\lambda>0.905$. For all viable Hu-Sawicki models, the Jeans limit is the same as for GR.

\subsection{Starobinsky model}


The Starobinsky model \cite{Starobinsky2007} is of the form
\be
f(R)=R-\lambda R_0\Big(\big(1+\frac{R^2}{R_0^2}\big)^{-n}-1\Big).
\end{equation}
This yields the condition $2n\lambda/R_0>1.3\times 10^{-31}\text{m}^{-2}$ for the globule FeSt 1-457. In \cite{Starobinsky2007} it is found that $n\ge0$ and $\lambda>8/3\sqrt3$. As the $R_0$ is of the order of the cosmological constant, {\it i.e.} very small, the condition is necessarily satisfied. The Starobinsky model exhibits assisted star formation due to the added $f(R)$ effects.

Both the Starobinsky model and the Hu-Sawicki model have similar expansions in the high curvature regime. These lead to a condition $n>0.9$ \cite{Capozziello:2007eu}. The shared condition is due to the similar expansion.

\subsection{Appleby-Battye model}
In the Appleby-Battye model the $f(R)$ is tailored to agree with the cosmology constraints as well as the stability issues. The form is
\be
f(R)=\frac 12R+\frac 1{2a}\log[\cosh(aR)-\tanh(b)\sinh(aR)],
\end{equation}
with $a$ and $b$ being the model parameters. This leads to $f''(R)=\frac a2\text{sech}^2(aR-b)$. In \cite{Appleby:2007vb}, it is found that $a\approx\frac{2b}{R_0}\approx\frac b{6H_0^2}$. Therefore
\be
\frac b{12H_0^2}\text{sech}^2(b)>1.3\times 10^{-31}\text{m}^{-2},
\end{equation}
which in turn implies that roughly $b<78$. Another constraint is that $8e^{-2b}\ll 2R_0$, which is satisfied around $b=46$, which leaves us a range of roughly $50<b<75$. The Appleby-Battye model requires fine-tuning due to the existing constraints. With the globule observations it is even more so.

\subsection{Tsujikawa model}
The Tsujikawa model is described by
\be
f(R)=R-\lambda R_c\tanh\Big(\frac R{R_c}\Big),
\ee
with $R_c$ and $\lambda$ being positive model parameters. The Tsujikawa and Appleby-Battye models have similarities, but for the purposes of our treatment, the behaviour is different.  
This model has $f''(R_0)=0$. Therefore, the Tsujikawa yields the same predictions for collapse in the globules as the conventional GR gravity.

\section{Conclusions}
We have examined the effects of $f(R)$ gravity on collisionless collapse, especially the limit of instability. The mass distribution is allowed to be time-dependent to better describe a collapse event. The examination is based on de Sitter background with perturbations.

We have found that with the addition of $f(R)$, the limit for collapse can be different. It is also found that with certain models, for which $f''(0)<0$ a new limit is present. This second limit is found to have no physical consequences and is ruled out due to the Dolgov-Kawasaki instability with present day curvatures.

In reference \cite{Capozziello:2011gm} an analysis similar to ours is done. However, the $f(R)$ parameter is fixed as $\alpha=-\frac{1}{k^2_j}=-\frac{\sigma^2}{4\pi G\rho_0}$ and the background is also taken to be Minkowskian. These are unnecessary constraints on the models, restricting to a fixed Jeans limit. Therefore, our results are more general. 

It is found that $f(R)$ models can affect star formation by lowering the limit for collapse. For viable models the result is assisted star formation. This is in agreement with the observed collapse behaviour in Bok globules. The globules experience star formation at rates higher than standard Jeans analysis would suggest.

The effects of $f(R)$ on Jeans mass are constrained for models passing the Dolgov-Kawasaki criterion. We find the lower limit for Jeans mass that a model could reach. The upper limit coincides with GR. In the extreme a modification can lower the required mass for collapse by around one third.

The modified Jeans limit, as well as the standard limit, are found to be of the order of Bok globules. These gas clouds and their collapse behaviour can be examined for agreement with $f(R)$ modified predictions.

We have used a small test sample of Bok globules to demonstrate that it is possible to obtain a constraint for some $f(R)$ models. This constraint is based on lowering the collapse to a level agreeing with the amount of protostars in Bok globules. With a larger data set and better understanding of the physics in these clouds, a strong limit might be obtained.

In our linearized approach, not all the characteristics of $f(R)$ models are present. It is also possible, that in a higher order, more theories would experience changes to stability. However, it is unlikely that these further changes would cancel the phenomena caused by the lower order terms.

Some of the examined viable $f(R)$ models revert to the standard GR value in regard of the modified Jeans limit. This is due to the modifications to the Jeans limit appearing only as $f''(R)\neq 0$. The Hu-Sawicki model and the Tsujikawa model do not experience any modifications as their acceptable parameter space does not allow for $f''(R_0)\neq 0$.

The Starobinsky model allows for the modified Jeans limit, which fit the observations well. For the Appleby-Battye model we find to obtain a considerably lower Jeans mass, which would better fit observations, the constraints on the model become even more stringent.

A more detailed collapse model, including {\it e.g.} turbulence, could provide a more accurate limit. Understanding the effects of modified gravity in star formation could lead to better understanding of the demands for a viable gravity theory.

The methodology we have developed in this article can be applied to more extensive datasets on Bok globules as they become available. Similar treatment can also be subjected to protogalaxies (in reference \cite{Roshan:2015gra} galactic disks are examined). Perhaps the most interesting possibility is to extend a similar treatment to other modified gravity theories such as scalar-tensor gravity. The effects on Jeans mass are likely to appear due to most modifications. Theories that raise Jeans mass inhibit star formation and therefore, are not favored by observations.

\section{Appendix}

\subsection{Dispersion relation integral}\label{append}
We examine the integral part of the dispersion relation
\begin{align}
&4 \pi G\int\Big(\frac{\mathbf k\cdot \frac{\partial f_0}{\partial\mathbf v}}{\mathbf v\cdot \mathbf k-\omega}\Big)d\mathbf v+\frac{k^4\Big(1+3\alpha(k^2-\omega^2+3\alpha\omega^4)\Big)}{(1+4\alpha k^2)(k^2-3\omega^2)} \nonumber\\
\equiv &I+\frac{k^4\Big(1+3\alpha(k^2-\omega^2+3\alpha\omega^4)\Big)}{(1+4\alpha k^2)(k^2-3\omega^2)}=0
\end{align}
The distribution of particle speeds in a stellar system follows the Maxwell-Boltzmann distribution. Therefore, we have for the background matter distribution $f_0(\mathbf v)$
\be
f_0=\frac{\rho_0}{\sqrt{2\pi\sigma^2}}e^{-(v^2/2\sigma^2)}.
\ee
The coordinate system is arbitrary and we are free to choose $\mathbf k=(k,0,0)$:
\be
I=-\frac{2\sqrt {2\pi} G \rho_0}{\sigma^3}\int\frac{kv_x e^{-v_x^2/(2\sigma^2)}}{kv_x-\omega}dv_x.
\ee
We make a substitution of $v_x=\sqrt 2\sigma x$ to reach
\be
-\frac{4\sqrt {\pi} G \rho_0}{\sigma^2}\int\frac{x e^{-x^2}}{x-\omega/(\sqrt 2\sigma k)}dx.
\ee
The problematic part is the singularity at $x=\omega/\sqrt 2\sigma k$. Depending on $\omega$, whether it is imaginary or not, the integration path must be chosen accordingly.  We are interested in the unstable modes for which $Im(\omega)>0$, which is also the most simple case. We notice that the integral has a close resemblance to a plasma dispersion function (\textit{e.g.} \cite{BT2} page 787)
\ba
Z(w)&=i\sqrt{\pi}e^{-w^2}\big(1+\text{erf}(\text{i} w)\big) , \ \ \ \big(\text{Im}(w)>0\big) \label{zwa}\\
&=\frac 1{\sqrt{\pi}}\int^\infty_{-\infty}\frac{e^{-s^2}}{s-w}ds
\end{align}
where $\text{erf}(z)$ is an error function. The $w$ derivative is found to be
\be
\frac{dZ(w)}{dw}=-\frac 2{\sqrt{\pi}}\int^\infty_{-\infty}\frac{se^{-s^2}}{s-w}ds
\ee
In general we have
\be
Z^{(n)}(w)=\frac{d^nZ(w)}{dw^n}=\frac{n!}{\sqrt n}\int^\infty_{-\infty}ds\frac{e^{-s^2}}{(s-w)^{n+1}}=\frac 1{\sqrt\pi}\int^\infty_{-\infty}\frac{ds}{s-w}\frac{d^n(e^{-s^2})}{ds^n}.
\ee
For Hermite polynomials $H_n(s)$ holds the equality
\be
\frac{d^n}{ds^n}(e^{-s^2})=(-1)^ne^{-s^2}H_n(s)
\ee
known as the Rodrigues formula. With this equality the derivatives of $Z(w)$ can be written as
\be
\frac{d^nZ(w)}{dw^n}=\frac{(-1)^n}{\sqrt n}\int^\infty_{-\infty}ds\frac{H_n(s)e^{-s^2}}{s-w}.
\ee
We can further use the Hermite polynomials in writing the powers of the variable $s$ with the relation
\be
s^n=\frac{1}{2^n}\sum^M_{m=0}d_m(n)H_{n-2m}(s)
\ee
with the coefficients $d_m(n)$ found in most tables and $M\equiv[n/2]$ and therefore
\be
Z_n(w)=\frac{1}{2^n}\sum^{[n/2]}_{m=0}(-1)^{n-2m}d_m(n)\frac{d^{n-2m}Z(w)}{dw^{n-2m}}.
\ee
This can be used to solve the integral in $I$
\be
\int\frac{x e^{-x^2}}{x-w}dx=1+wZ(w).
\ee
The imaginary part of $wZ(w)$ must vanish for the dispersion relation to be satisfied. For that to happen, we must have $\text{Re}(w)=0$. We further mark $\omega=i\omega_I$ and write the plasma dispersion function in a different form \eqref{zwa}
\be
1+wZ(w)=1+\text{i}w\sqrt\pi e^{-w^2}\big[1+\text{erf}(\text{i}w)\big]=1-\frac{\sqrt\pi\omega_I}{\sqrt 2k\sigma}\text{exp}\Big(\frac{\omega_I^2}{2k^2\sigma^2}\Big)\text{erfc}\Big(\frac{\omega_I}{\sqrt 2k\sigma}\Big)
\ee
with $\text{erfc}(z)\equiv 1-\text{erf}(z)$ being the complementary error function and $w=\omega/\sqrt{2}k\sigma$, $\text{erf}(-z)=-\text{erf}(z)$. Finally we have
\be
I=-\frac{4\pi G \rho_0}{\sigma^2}\Big[1-\frac{\sqrt\pi\omega_I}{\sqrt 2k\sigma}\text{exp}\Big(\frac{\omega_I^2}{2k^2\sigma^2}\Big)\text{erfc}\big(\frac{\omega_I}{\sqrt 2k\sigma}\big)\Big].
\ee

\bibliographystyle{apsrev4-1}
\bibliography{refs_thesis}

\begin{thebibliography}{57}%
\makeatletter
\providecommand \@ifxundefined [1]{%
 \@ifx{#1\undefined}
}%
\providecommand \@ifnum [1]{%
 \ifnum #1\expandafter \@firstoftwo
 \else \expandafter \@secondoftwo
 \fi
}%
\providecommand \@ifx [1]{%
 \ifx #1\expandafter \@firstoftwo
 \else \expandafter \@secondoftwo
 \fi
}%
\providecommand \natexlab [1]{#1}%
\providecommand \enquote  [1]{``#1''}%
\providecommand \bibnamefont  [1]{#1}%
\providecommand \bibfnamefont [1]{#1}%
\providecommand \citenamefont [1]{#1}%
\providecommand \href@noop [0]{\@secondoftwo}%
\providecommand \href [0]{\begingroup \@sanitize@url \@href}%
\providecommand \@href[1]{\@@startlink{#1}\@@href}%
\providecommand \@@href[1]{\endgroup#1\@@endlink}%
\providecommand \@sanitize@url [0]{\catcode `\\12\catcode `\$12\catcode
  `\&12\catcode `\#12\catcode `\^12\catcode `\_12\catcode `\%12\relax}%
\providecommand \@@startlink[1]{}%
\providecommand \@@endlink[0]{}%
\providecommand \url  [0]{\begingroup\@sanitize@url \@url }%
\providecommand \@url [1]{\endgroup\@href {#1}{\urlprefix }}%
\providecommand \urlprefix  [0]{URL }%
\providecommand \Eprint [0]{\href }%
\providecommand \doibase [0]{http://dx.doi.org/}%
\providecommand \selectlanguage [0]{\@gobble}%
\providecommand \bibinfo  [0]{\@secondoftwo}%
\providecommand \bibfield  [0]{\@secondoftwo}%
\providecommand \translation [1]{[#1]}%
\providecommand \BibitemOpen [0]{}%
\providecommand \bibitemStop [0]{}%
\providecommand \bibitemNoStop [0]{.\EOS\space}%
\providecommand \EOS [0]{\spacefactor3000\relax}%
\providecommand \BibitemShut  [1]{\csname bibitem#1\endcsname}%
\let\auto@bib@innerbib\@empty
\bibitem [{\citenamefont {Wang}\ \emph {et~al.}(2000)\citenamefont {Wang},
  \citenamefont {Caldwell}, \citenamefont {Ostriker},\ and\ \citenamefont
  {Steinhardt}}]{Wang:1999fa}%
  \BibitemOpen
  \bibfield  {author} {\bibinfo {author} {\bibfnamefont {L.-M.}\ \bibnamefont
  {Wang}}, \bibinfo {author} {\bibfnamefont {R.~R.}\ \bibnamefont {Caldwell}},
  \bibinfo {author} {\bibfnamefont {J.~P.}\ \bibnamefont {Ostriker}}, \ and\
  \bibinfo {author} {\bibfnamefont {P.~J.}\ \bibnamefont {Steinhardt}},\ }\href
  {\doibase 10.1086/308331} {\bibfield  {journal} {\bibinfo  {journal}
  {Astrophys. J.}\ }\textbf {\bibinfo {volume} {530}},\ \bibinfo {pages} {17}
  (\bibinfo {year} {2000})},\ \Eprint {http://arxiv.org/abs/astro-ph/9901388}
  {arXiv:astro-ph/9901388 [astro-ph]} \BibitemShut {NoStop}%
\bibitem [{\citenamefont {Perlmutter}\ \emph {et~al.}(1999)\citenamefont
  {Perlmutter}, \citenamefont {Turner},\ and\ \citenamefont
  {White}}]{Perlmutter1999}%
  \BibitemOpen
  \bibfield  {author} {\bibinfo {author} {\bibfnamefont {S.}~\bibnamefont
  {Perlmutter}}, \bibinfo {author} {\bibfnamefont {M.~S.}\ \bibnamefont
  {Turner}}, \ and\ \bibinfo {author} {\bibfnamefont {M.~J.}\ \bibnamefont
  {White}},\ }\href {\doibase 10.1103/PhysRevLett.83.670} {\bibfield  {journal}
  {\bibinfo  {journal} {Phys. Rev. Lett.}\ }\textbf {\bibinfo {volume} {83}},\
  \bibinfo {pages} {670} (\bibinfo {year} {1999})},\ \Eprint
  {http://arxiv.org/abs/astro-ph/9901052} {arXiv:astro-ph/9901052} \BibitemShut
  {NoStop}%
\bibitem [{\citenamefont {Riess}\ \emph {et~al.}(1998)\citenamefont {Riess}
  \emph {et~al.}}]{riess98}%
  \BibitemOpen
  \bibfield  {author} {\bibinfo {author} {\bibfnamefont {A.~G.}\ \bibnamefont
  {Riess}} \emph {et~al.} (\bibinfo {collaboration} {Supernova Search Team}),\
  }\href {\doibase 10.1086/300499} {\bibfield  {journal} {\bibinfo  {journal}
  {Astron. J.}\ }\textbf {\bibinfo {volume} {116}},\ \bibinfo {pages} {1009}
  (\bibinfo {year} {1998})},\ \Eprint {http://arxiv.org/abs/astro-ph/9805201}
  {arXiv:astro-ph/9805201} \BibitemShut {NoStop}%
\bibitem [{\citenamefont {Weinberg}(1989)}]{Weinberg:1988cp}%
  \BibitemOpen
  \bibfield  {author} {\bibinfo {author} {\bibfnamefont {S.}~\bibnamefont
  {Weinberg}},\ }\href {\doibase 10.1103/RevModPhys.61.1} {\bibfield  {journal}
  {\bibinfo  {journal} {Rev. Mod. Phys.}\ }\textbf {\bibinfo {volume} {61}},\
  \bibinfo {pages} {1} (\bibinfo {year} {1989})}\BibitemShut {NoStop}%
\bibitem [{\citenamefont {Bili?}\ \emph {et~al.}(2002)\citenamefont {Bili?},
  \citenamefont {Tupper},\ and\ \citenamefont {Viollier}}]{Bilic200217}%
  \BibitemOpen
  \bibfield  {author} {\bibinfo {author} {\bibfnamefont {N.}~\bibnamefont
  {Bili?}}, \bibinfo {author} {\bibfnamefont {G.~B.}\ \bibnamefont {Tupper}}, \
  and\ \bibinfo {author} {\bibfnamefont {R.~D.}\ \bibnamefont {Viollier}},\
  }\href {\doibase http://dx.doi.org/10.1016/S0370-2693(02)01716-1} {\bibfield
  {journal} {\bibinfo  {journal} {Physics Letters B}\ }\textbf {\bibinfo
  {volume} {535}},\ \bibinfo {pages} {17 } (\bibinfo {year}
  {2002})}\BibitemShut {NoStop}%
\bibitem [{\citenamefont {Bento}\ \emph {et~al.}(2002)\citenamefont {Bento},
  \citenamefont {Bertolami},\ and\ \citenamefont {Sen}}]{Bento:2002ps}%
  \BibitemOpen
  \bibfield  {author} {\bibinfo {author} {\bibfnamefont {M.~C.}\ \bibnamefont
  {Bento}}, \bibinfo {author} {\bibfnamefont {O.}~\bibnamefont {Bertolami}}, \
  and\ \bibinfo {author} {\bibfnamefont {A.~A.}\ \bibnamefont {Sen}},\ }\href
  {\doibase 10.1103/PhysRevD.66.043507} {\bibfield  {journal} {\bibinfo
  {journal} {Phys. Rev.}\ }\textbf {\bibinfo {volume} {D66}},\ \bibinfo {pages}
  {043507} (\bibinfo {year} {2002})},\ \Eprint
  {http://arxiv.org/abs/gr-qc/0202064} {arXiv:gr-qc/0202064 [gr-qc]}
  \BibitemShut {NoStop}%
\bibitem [{\citenamefont {Bolejko}\ \emph {et~al.}(2011)\citenamefont
  {Bolejko}, \citenamefont {Celerier},\ and\ \citenamefont
  {Krasinski}}]{Bolejko:2011jc}%
  \BibitemOpen
  \bibfield  {author} {\bibinfo {author} {\bibfnamefont {K.}~\bibnamefont
  {Bolejko}}, \bibinfo {author} {\bibfnamefont {M.-N.}\ \bibnamefont
  {Celerier}}, \ and\ \bibinfo {author} {\bibfnamefont {A.}~\bibnamefont
  {Krasinski}},\ }\href {\doibase 10.1088/0264-9381/28/16/164002} {\bibfield
  {journal} {\bibinfo  {journal} {Class. Quant. Grav.}\ }\textbf {\bibinfo
  {volume} {28}},\ \bibinfo {pages} {164002} (\bibinfo {year} {2011})},\
  \Eprint {http://arxiv.org/abs/1102.1449} {arXiv:1102.1449 [astro-ph.CO]}
  \BibitemShut {NoStop}%
\bibitem [{\citenamefont {Clifton}\ \emph {et~al.}(2012)\citenamefont
  {Clifton}, \citenamefont {Ferreira}, \citenamefont {Padilla},\ and\
  \citenamefont {Skordis}}]{Clifton:2011jh}%
  \BibitemOpen
  \bibfield  {author} {\bibinfo {author} {\bibfnamefont {T.}~\bibnamefont
  {Clifton}}, \bibinfo {author} {\bibfnamefont {P.~G.}\ \bibnamefont
  {Ferreira}}, \bibinfo {author} {\bibfnamefont {A.}~\bibnamefont {Padilla}}, \
  and\ \bibinfo {author} {\bibfnamefont {C.}~\bibnamefont {Skordis}},\ }\href
  {\doibase 10.1016/j.physrep.2012.01.001} {\bibfield  {journal} {\bibinfo
  {journal} {Phys. Rept.}\ }\textbf {\bibinfo {volume} {513}},\ \bibinfo
  {pages} {1} (\bibinfo {year} {2012})},\ \Eprint
  {http://arxiv.org/abs/1106.2476} {arXiv:1106.2476 [astro-ph.CO]} \BibitemShut
  {NoStop}%
\bibitem [{\citenamefont {Flanagan}(2004)}]{flanagan2003}%
  \BibitemOpen
  \bibfield  {author} {\bibinfo {author} {\bibfnamefont {E.~E.}\ \bibnamefont
  {Flanagan}},\ }\href@noop {} {\bibfield  {journal} {\bibinfo  {journal}
  {Phys. Rev. Lett.}\ }\textbf {\bibinfo {volume} {92}},\ \bibinfo {pages}
  {071101} (\bibinfo {year} {2004})},\ \Eprint
  {http://arxiv.org/abs/astro-ph/0308111} {astro-ph/0308111} \BibitemShut
  {NoStop}%
\bibitem [{\citenamefont {Hu}\ and\ \citenamefont {Sawicki}(2007)}]{Hu2007}%
  \BibitemOpen
  \bibfield  {author} {\bibinfo {author} {\bibfnamefont {W.}~\bibnamefont
  {Hu}}\ and\ \bibinfo {author} {\bibfnamefont {I.}~\bibnamefont {Sawicki}},\
  }\href {\doibase 10.1103/PhysRevD.76.064004} {\bibfield  {journal} {\bibinfo
  {journal} {Phys. Rev.}\ }\textbf {\bibinfo {volume} {D76}},\ \bibinfo {pages}
  {064004} (\bibinfo {year} {2007})},\ \Eprint {http://arxiv.org/abs/0705.1158}
  {arXiv:0705.1158 [astro-ph]} \BibitemShut {NoStop}%
\bibitem [{\citenamefont {Starobinsky}(2007)}]{Starobinsky2007}%
  \BibitemOpen
  \bibfield  {author} {\bibinfo {author} {\bibfnamefont {A.~A.}\ \bibnamefont
  {Starobinsky}},\ }\href {\doibase 10.1134/S0021364007150027} {\bibfield
  {journal} {\bibinfo  {journal} {JETP Lett.}\ }\textbf {\bibinfo {volume}
  {86}},\ \bibinfo {pages} {157} (\bibinfo {year} {2007})},\ \Eprint
  {http://arxiv.org/abs/0706.2041} {arXiv:0706.2041 [astro-ph]} \BibitemShut
  {NoStop}%
\bibitem [{\citenamefont {Tsujikawa}(2008)}]{Tsujikawa2007}%
  \BibitemOpen
  \bibfield  {author} {\bibinfo {author} {\bibfnamefont {S.}~\bibnamefont
  {Tsujikawa}},\ }\href {\doibase 10.1103/PhysRevD.77.023507} {\bibfield
  {journal} {\bibinfo  {journal} {Phys. Rev.}\ }\textbf {\bibinfo {volume}
  {D77}},\ \bibinfo {pages} {023507} (\bibinfo {year} {2008})},\ \Eprint
  {http://arxiv.org/abs/0709.1391} {arXiv:0709.1391 [astro-ph]} \BibitemShut
  {NoStop}%
\bibitem [{\citenamefont {Appleby}\ and\ \citenamefont
  {Battye}(2007)}]{Appleby:2007vb}%
  \BibitemOpen
  \bibfield  {author} {\bibinfo {author} {\bibfnamefont {S.~A.}\ \bibnamefont
  {Appleby}}\ and\ \bibinfo {author} {\bibfnamefont {R.~A.}\ \bibnamefont
  {Battye}},\ }\href {\doibase 10.1016/j.physletb.2007.08.037} {\bibfield
  {journal} {\bibinfo  {journal} {Phys.Lett.}\ }\textbf {\bibinfo {volume}
  {B654}},\ \bibinfo {pages} {7} (\bibinfo {year} {2007})},\ \Eprint
  {http://arxiv.org/abs/0705.3199} {arXiv:0705.3199 [astro-ph]} \BibitemShut
  {NoStop}%
\bibitem [{\citenamefont {Dolgov}\ and\ \citenamefont
  {Kawasaki}(2003)}]{Dolgov2003}%
  \BibitemOpen
  \bibfield  {author} {\bibinfo {author} {\bibfnamefont {A.~D.}\ \bibnamefont
  {Dolgov}}\ and\ \bibinfo {author} {\bibfnamefont {M.}~\bibnamefont
  {Kawasaki}},\ }\href {\doibase 10.1016/j.physletb.2003.08.039} {\bibfield
  {journal} {\bibinfo  {journal} {Phys. Lett.}\ }\textbf {\bibinfo {volume}
  {B573}},\ \bibinfo {pages} {1} (\bibinfo {year} {2003})},\ \Eprint
  {http://arxiv.org/abs/astro-ph/0307285} {arXiv:astro-ph/0307285} \BibitemShut
  {NoStop}%
\bibitem [{\citenamefont {Chiba}(2003)}]{Chiba2003}%
  \BibitemOpen
  \bibfield  {author} {\bibinfo {author} {\bibfnamefont {T.}~\bibnamefont
  {Chiba}},\ }\href@noop {} {\bibfield  {journal} {\bibinfo  {journal} {Phys.
  Lett.}\ }\textbf {\bibinfo {volume} {B575}},\ \bibinfo {pages} {1} (\bibinfo
  {year} {2003})},\ \Eprint {http://arxiv.org/abs/astro-ph/0307338}
  {astro-ph/0307338} \BibitemShut {NoStop}%
\bibitem [{\citenamefont {Khoury}\ and\ \citenamefont
  {Weltman}(2004)}]{Khoury:2003rn}%
  \BibitemOpen
  \bibfield  {author} {\bibinfo {author} {\bibfnamefont {J.}~\bibnamefont
  {Khoury}}\ and\ \bibinfo {author} {\bibfnamefont {A.}~\bibnamefont
  {Weltman}},\ }\href {\doibase 10.1103/PhysRevD.69.044026} {\bibfield
  {journal} {\bibinfo  {journal} {Phys.Rev.}\ }\textbf {\bibinfo {volume}
  {D69}},\ \bibinfo {pages} {044026} (\bibinfo {year} {2004})},\ \Eprint
  {http://arxiv.org/abs/astro-ph/0309411} {arXiv:astro-ph/0309411 [astro-ph]}
  \BibitemShut {NoStop}%
\bibitem [{\citenamefont {Song}\ \emph {et~al.}(2007)\citenamefont {Song},
  \citenamefont {Hu},\ and\ \citenamefont {Sawicki}}]{Song:2006ej}%
  \BibitemOpen
  \bibfield  {author} {\bibinfo {author} {\bibfnamefont {Y.-S.}\ \bibnamefont
  {Song}}, \bibinfo {author} {\bibfnamefont {W.}~\bibnamefont {Hu}}, \ and\
  \bibinfo {author} {\bibfnamefont {I.}~\bibnamefont {Sawicki}},\ }\href
  {\doibase 10.1103/PhysRevD.75.044004} {\bibfield  {journal} {\bibinfo
  {journal} {Phys.Rev.}\ }\textbf {\bibinfo {volume} {D75}},\ \bibinfo {pages}
  {044004} (\bibinfo {year} {2007})},\ \Eprint
  {http://arxiv.org/abs/astro-ph/0610532} {arXiv:astro-ph/0610532 [astro-ph]}
  \BibitemShut {NoStop}%
\bibitem [{\citenamefont {Amendola}\ \emph {et~al.}(2007)\citenamefont
  {Amendola}, \citenamefont {Polarski},\ and\ \citenamefont
  {Tsujikawa}}]{Amendola:2006kh}%
  \BibitemOpen
  \bibfield  {author} {\bibinfo {author} {\bibfnamefont {L.}~\bibnamefont
  {Amendola}}, \bibinfo {author} {\bibfnamefont {D.}~\bibnamefont {Polarski}},
  \ and\ \bibinfo {author} {\bibfnamefont {S.}~\bibnamefont {Tsujikawa}},\
  }\href {\doibase 10.1103/PhysRevLett.98.131302} {\bibfield  {journal}
  {\bibinfo  {journal} {Phys.Rev.Lett.}\ }\textbf {\bibinfo {volume} {98}},\
  \bibinfo {pages} {131302} (\bibinfo {year} {2007})},\ \Eprint
  {http://arxiv.org/abs/astro-ph/0603703} {arXiv:astro-ph/0603703 [astro-ph]}
  \BibitemShut {NoStop}%
\bibitem [{\citenamefont {Nojiri}\ and\ \citenamefont
  {Odintsov}(2008)}]{PhysRevD.77.026007}%
  \BibitemOpen
  \bibfield  {author} {\bibinfo {author} {\bibfnamefont {S.}~\bibnamefont
  {Nojiri}}\ and\ \bibinfo {author} {\bibfnamefont {S.~D.}\ \bibnamefont
  {Odintsov}},\ }\href {\doibase 10.1103/PhysRevD.77.026007} {\bibfield
  {journal} {\bibinfo  {journal} {Phys. Rev. D}\ }\textbf {\bibinfo {volume}
  {77}},\ \bibinfo {pages} {026007} (\bibinfo {year} {2008})}\BibitemShut
  {NoStop}%
\bibitem [{\citenamefont {Boehmer}\ \emph {et~al.}(2008)\citenamefont
  {Boehmer}, \citenamefont {Harko},\ and\ \citenamefont
  {Lobo}}]{Boehmer:2007kx}%
  \BibitemOpen
  \bibfield  {author} {\bibinfo {author} {\bibfnamefont {C.~G.}\ \bibnamefont
  {Boehmer}}, \bibinfo {author} {\bibfnamefont {T.}~\bibnamefont {Harko}}, \
  and\ \bibinfo {author} {\bibfnamefont {F.~S.~N.}\ \bibnamefont {Lobo}},\
  }\href {\doibase 10.1016/j.astropartphys.2008.04.003} {\bibfield  {journal}
  {\bibinfo  {journal} {Astropart. Phys.}\ }\textbf {\bibinfo {volume} {29}},\
  \bibinfo {pages} {386} (\bibinfo {year} {2008})},\ \Eprint
  {http://arxiv.org/abs/0709.0046} {arXiv:0709.0046 [gr-qc]} \BibitemShut
  {NoStop}%
\bibitem [{\citenamefont {Capozziello}\ \emph
  {et~al.}(2007{\natexlab{a}})\citenamefont {Capozziello}, \citenamefont
  {Cardone},\ and\ \citenamefont {Troisi}}]{Capozziello:2006ph}%
  \BibitemOpen
  \bibfield  {author} {\bibinfo {author} {\bibfnamefont {S.}~\bibnamefont
  {Capozziello}}, \bibinfo {author} {\bibfnamefont {V.}~\bibnamefont
  {Cardone}}, \ and\ \bibinfo {author} {\bibfnamefont {A.}~\bibnamefont
  {Troisi}},\ }\href {\doibase 10.1111/j.1365-2966.2007.11401.x} {\bibfield
  {journal} {\bibinfo  {journal} {Mon.Not.Roy.Astron.Soc.}\ }\textbf {\bibinfo
  {volume} {375}},\ \bibinfo {pages} {1423} (\bibinfo {year}
  {2007}{\natexlab{a}})},\ \Eprint {http://arxiv.org/abs/astro-ph/0603522}
  {arXiv:astro-ph/0603522 [astro-ph]} \BibitemShut {NoStop}%
\bibitem [{\citenamefont {Capozziello}\ \emph {et~al.}(2009)\citenamefont
  {Capozziello}, \citenamefont {De~Filippis},\ and\ \citenamefont
  {Salzano}}]{Capozziello:2008ny}%
  \BibitemOpen
  \bibfield  {author} {\bibinfo {author} {\bibfnamefont {S.}~\bibnamefont
  {Capozziello}}, \bibinfo {author} {\bibfnamefont {E.}~\bibnamefont
  {De~Filippis}}, \ and\ \bibinfo {author} {\bibfnamefont {V.}~\bibnamefont
  {Salzano}},\ }\href {\doibase 10.1111/j.1365-2966.2008.14382.x} {\bibfield
  {journal} {\bibinfo  {journal} {Mon.Not.Roy.Astron.Soc.}\ }\textbf {\bibinfo
  {volume} {394}},\ \bibinfo {pages} {947} (\bibinfo {year} {2009})},\ \Eprint
  {http://arxiv.org/abs/0809.1882} {arXiv:0809.1882 [astro-ph]} \BibitemShut
  {NoStop}%
\bibitem [{\citenamefont {Cooney}\ \emph {et~al.}(2010)\citenamefont {Cooney},
  \citenamefont {DeDeo},\ and\ \citenamefont {Psaltis}}]{Cooney:2009rr}%
  \BibitemOpen
  \bibfield  {author} {\bibinfo {author} {\bibfnamefont {A.}~\bibnamefont
  {Cooney}}, \bibinfo {author} {\bibfnamefont {S.}~\bibnamefont {DeDeo}}, \
  and\ \bibinfo {author} {\bibfnamefont {D.}~\bibnamefont {Psaltis}},\ }\href
  {\doibase 10.1103/PhysRevD.82.064033} {\bibfield  {journal} {\bibinfo
  {journal} {Phys.Rev.}\ }\textbf {\bibinfo {volume} {D82}},\ \bibinfo {pages}
  {064033} (\bibinfo {year} {2010})},\ \Eprint {http://arxiv.org/abs/0910.5480}
  {arXiv:0910.5480 [astro-ph.HE]} \BibitemShut {NoStop}%
\bibitem [{\citenamefont {Jeans}(1902)}]{jeans_original}%
  \BibitemOpen
  \bibfield  {author} {\bibinfo {author} {\bibfnamefont {J.~H.}\ \bibnamefont
  {Jeans}},\ }\href {\doibase 10.1098/rsta.1902.0012} {\bibfield  {journal}
  {\bibinfo  {journal} {Philosophical Transactions of the Royal Society of
  London A: Mathematical, Physical and Engineering Sciences}\ }\textbf
  {\bibinfo {volume} {199}},\ \bibinfo {pages} {1} (\bibinfo {year}
  {1902})}\BibitemShut {NoStop}%
\bibitem [{\citenamefont {Capozziello}\ \emph {et~al.}(2012)\citenamefont
  {Capozziello}, \citenamefont {De~Laurentis}, \citenamefont {De~Martino},
  \citenamefont {Formisano},\ and\ \citenamefont
  {Odintsov}}]{Capozziello:2011gm}%
  \BibitemOpen
  \bibfield  {author} {\bibinfo {author} {\bibfnamefont {S.}~\bibnamefont
  {Capozziello}}, \bibinfo {author} {\bibfnamefont {M.}~\bibnamefont
  {De~Laurentis}}, \bibinfo {author} {\bibfnamefont {I.}~\bibnamefont
  {De~Martino}}, \bibinfo {author} {\bibfnamefont {M.}~\bibnamefont
  {Formisano}}, \ and\ \bibinfo {author} {\bibfnamefont {S.}~\bibnamefont
  {Odintsov}},\ }\href {\doibase 10.1103/PhysRevD.85.044022} {\bibfield
  {journal} {\bibinfo  {journal} {Phys.Rev.}\ }\textbf {\bibinfo {volume}
  {D85}},\ \bibinfo {pages} {044022} (\bibinfo {year} {2012})},\ \Eprint
  {http://arxiv.org/abs/1112.0761} {arXiv:1112.0761 [gr-qc]} \BibitemShut
  {NoStop}%
\bibitem [{\citenamefont {Roshan}\ and\ \citenamefont
  {Abbassi}(2014)}]{Roshan:2014mqa}%
  \BibitemOpen
  \bibfield  {author} {\bibinfo {author} {\bibfnamefont {M.}~\bibnamefont
  {Roshan}}\ and\ \bibinfo {author} {\bibfnamefont {S.}~\bibnamefont
  {Abbassi}},\ }\href {\doibase 10.1103/PhysRevD.90.044010} {\bibfield
  {journal} {\bibinfo  {journal} {Phys.Rev.}\ }\textbf {\bibinfo {volume}
  {D90}},\ \bibinfo {pages} {044010} (\bibinfo {year} {2014})},\ \Eprint
  {http://arxiv.org/abs/1407.6431} {arXiv:1407.6431 [astro-ph.GA]} \BibitemShut
  {NoStop}%
\bibitem [{\citenamefont {Roman-Duval}\ \emph {et~al.}(2010)\citenamefont
  {Roman-Duval}, \citenamefont {Jackson}, \citenamefont {Heyer}, \citenamefont
  {Rathborne},\ and\ \citenamefont {Simon}}]{RomanDuval:2010nm}%
  \BibitemOpen
  \bibfield  {author} {\bibinfo {author} {\bibfnamefont {J.}~\bibnamefont
  {Roman-Duval}}, \bibinfo {author} {\bibfnamefont {J.~M.}\ \bibnamefont
  {Jackson}}, \bibinfo {author} {\bibfnamefont {M.}~\bibnamefont {Heyer}},
  \bibinfo {author} {\bibfnamefont {J.}~\bibnamefont {Rathborne}}, \ and\
  \bibinfo {author} {\bibfnamefont {R.}~\bibnamefont {Simon}},\ }\href
  {\doibase 10.1088/0004-637X/723/1/492} {\bibfield  {journal} {\bibinfo
  {journal} {Astrophys.J.}\ }\textbf {\bibinfo {volume} {723}},\ \bibinfo
  {pages} {492} (\bibinfo {year} {2010})},\ \Eprint
  {http://arxiv.org/abs/1010.2798} {arXiv:1010.2798 [astro-ph.GA]} \BibitemShut
  {NoStop}%
\bibitem [{\citenamefont {Mukhanov}\ \emph {et~al.}(1992)\citenamefont
  {Mukhanov}, \citenamefont {Feldman},\ and\ \citenamefont
  {Brandenberger}}]{mukhanov1990}%
  \BibitemOpen
  \bibfield  {author} {\bibinfo {author} {\bibfnamefont {V.~F.}\ \bibnamefont
  {Mukhanov}}, \bibinfo {author} {\bibfnamefont {H.~A.}\ \bibnamefont
  {Feldman}}, \ and\ \bibinfo {author} {\bibfnamefont {R.~H.}\ \bibnamefont
  {Brandenberger}},\ }\href {\doibase 10.1016/0370-1573(92)90044-Z} {\bibfield
  {journal} {\bibinfo  {journal} {Phys. Rept.}\ }\textbf {\bibinfo {volume}
  {215}},\ \bibinfo {pages} {203} (\bibinfo {year} {1992})}\BibitemShut
  {NoStop}%
\bibitem [{\citenamefont {Weinberg}(1972)}]{weinberg1972}%
  \BibitemOpen
  \bibfield  {author} {\bibinfo {author} {\bibfnamefont {S.}~\bibnamefont
  {Weinberg}},\ }\href@noop {} {\emph {\bibinfo {title} {Gravitation and
  Cosmology: Principles and Applications of the General Theory of
  Relativity}}}\ (\bibinfo  {publisher} {John Wiley et Sons, Inc.},\ \bibinfo
  {address} {New York, NY, USA},\ \bibinfo {year} {1972})\BibitemShut {NoStop}%
\bibitem [{\citenamefont {Jain}\ \emph {et~al.}(2013)\citenamefont {Jain},
  \citenamefont {Vikram},\ and\ \citenamefont {Sakstein}}]{Jain:2012tn}%
  \BibitemOpen
  \bibfield  {author} {\bibinfo {author} {\bibfnamefont {B.}~\bibnamefont
  {Jain}}, \bibinfo {author} {\bibfnamefont {V.}~\bibnamefont {Vikram}}, \ and\
  \bibinfo {author} {\bibfnamefont {J.}~\bibnamefont {Sakstein}},\ }\href
  {\doibase 10.1088/0004-637X/779/1/39} {\bibfield  {journal} {\bibinfo
  {journal} {Astrophys. J.}\ }\textbf {\bibinfo {volume} {779}},\ \bibinfo
  {pages} {39} (\bibinfo {year} {2013})},\ \Eprint
  {http://arxiv.org/abs/1204.6044} {arXiv:1204.6044 [astro-ph.CO]} \BibitemShut
  {NoStop}%
\bibitem [{\citenamefont {Ade}\ \emph {et~al.}(2015)\citenamefont {Ade} \emph
  {et~al.}}]{Ade:2015xua}%
  \BibitemOpen
  \bibfield  {author} {\bibinfo {author} {\bibfnamefont {P.~A.~R.}\
  \bibnamefont {Ade}} \emph {et~al.} (\bibinfo {collaboration} {Planck}),\
  }\href@noop {} {\  (\bibinfo {year} {2015})},\ \Eprint
  {http://arxiv.org/abs/1502.01589} {arXiv:1502.01589 [astro-ph.CO]}
  \BibitemShut {NoStop}%
\bibitem [{\citenamefont {Capozziello}\ and\ \citenamefont
  {Saez-Gomez}(2012)}]{Capozziello:2011wg}%
  \BibitemOpen
  \bibfield  {author} {\bibinfo {author} {\bibfnamefont {S.}~\bibnamefont
  {Capozziello}}\ and\ \bibinfo {author} {\bibfnamefont {D.}~\bibnamefont
  {Saez-Gomez}},\ }\href {\doibase 10.1002/andp.201100244} {\bibfield
  {journal} {\bibinfo  {journal} {Annalen Phys.}\ }\textbf {\bibinfo {volume}
  {524}},\ \bibinfo {pages} {279} (\bibinfo {year} {2012})},\ \Eprint
  {http://arxiv.org/abs/1107.0948} {arXiv:1107.0948 [gr-qc]} \BibitemShut
  {NoStop}%
\bibitem [{\citenamefont {Olive}\ \emph {et~al.}(2014)\citenamefont {Olive}
  \emph {et~al.}}]{Agashe:2014kda}%
  \BibitemOpen
  \bibfield  {author} {\bibinfo {author} {\bibfnamefont {K.~A.}\ \bibnamefont
  {Olive}} \emph {et~al.} (\bibinfo {collaboration} {Particle Data Group}),\
  }\href {\doibase 10.1088/1674-1137/38/9/090001} {\bibfield  {journal}
  {\bibinfo  {journal} {Chin. Phys.}\ }\textbf {\bibinfo {volume} {C38}},\
  \bibinfo {pages} {090001} (\bibinfo {year} {2014})}\BibitemShut {NoStop}%
\bibitem [{\citenamefont {Capozziello}\ \emph
  {et~al.}(2007{\natexlab{b}})\citenamefont {Capozziello}, \citenamefont
  {Stabile},\ and\ \citenamefont {Troisi}}]{Capozziello:2007ms}%
  \BibitemOpen
  \bibfield  {author} {\bibinfo {author} {\bibfnamefont {S.}~\bibnamefont
  {Capozziello}}, \bibinfo {author} {\bibfnamefont {A.}~\bibnamefont
  {Stabile}}, \ and\ \bibinfo {author} {\bibfnamefont {A.}~\bibnamefont
  {Troisi}},\ }\href {\doibase 10.1103/PhysRevD.76.104019} {\bibfield
  {journal} {\bibinfo  {journal} {Phys. Rev.}\ }\textbf {\bibinfo {volume}
  {D76}},\ \bibinfo {pages} {104019} (\bibinfo {year} {2007}{\natexlab{b}})},\
  \Eprint {http://arxiv.org/abs/0708.0723} {arXiv:0708.0723 [gr-qc]}
  \BibitemShut {NoStop}%
\bibitem [{\citenamefont {Binney}\ and\ \citenamefont {Tremaine}(2008)}]{BT2}%
  \BibitemOpen
  \bibfield  {author} {\bibinfo {author} {\bibfnamefont {J.}~\bibnamefont
  {Binney}}\ and\ \bibinfo {author} {\bibfnamefont {S.}~\bibnamefont
  {Tremaine}},\ }\href@noop {} {\emph {\bibinfo {title} {Galactic Dynamics}}},\
  \bibinfo {edition} {2nd}\ ed.,\ Princeton Series in Astrophysics\ (\bibinfo
  {publisher} {Princeton University Press},\ \bibinfo {year}
  {2008})\BibitemShut {NoStop}%
\bibitem [{\citenamefont {Faraoni}(2004)}]{faraoni04}%
  \BibitemOpen
  \bibfield  {author} {\bibinfo {author} {\bibfnamefont {V.}~\bibnamefont
  {Faraoni}},\ }\href@noop {} {\emph {\bibinfo {title} {Cosmology in
  Scalar-Tensor Gravity}}}\ (\bibinfo  {publisher} {Kluwer Academic Publishers
  Group},\ \bibinfo {address} {Norwell, MA, USA, and Dordrecht, The
  Netherlands},\ \bibinfo {year} {2004})\BibitemShut {NoStop}%
\bibitem [{\citenamefont {Sotiriou}\ and\ \citenamefont
  {Faraoni}(2010)}]{Sotiriou:2008rp}%
  \BibitemOpen
  \bibfield  {author} {\bibinfo {author} {\bibfnamefont {T.~P.}\ \bibnamefont
  {Sotiriou}}\ and\ \bibinfo {author} {\bibfnamefont {V.}~\bibnamefont
  {Faraoni}},\ }\href {\doibase 10.1103/RevModPhys.82.451} {\bibfield
  {journal} {\bibinfo  {journal} {Rev. Mod. Phys.}\ }\textbf {\bibinfo {volume}
  {82}},\ \bibinfo {pages} {451} (\bibinfo {year} {2010})},\ \Eprint
  {http://arxiv.org/abs/0805.1726} {arXiv:0805.1726 [gr-qc]} \BibitemShut
  {NoStop}%
\bibitem [{\citenamefont {Arbuzova}\ \emph {et~al.}(2015)\citenamefont
  {Arbuzova}, \citenamefont {Dolgov},\ and\ \citenamefont
  {Reverberi}}]{Arbuzova:2015uga}%
  \BibitemOpen
  \bibfield  {author} {\bibinfo {author} {\bibfnamefont {E.~V.}\ \bibnamefont
  {Arbuzova}}, \bibinfo {author} {\bibfnamefont {A.~D.}\ \bibnamefont
  {Dolgov}}, \ and\ \bibinfo {author} {\bibfnamefont {L.}~\bibnamefont
  {Reverberi}},\ }\href {\doibase 10.1103/PhysRevD.92.064041} {\bibfield
  {journal} {\bibinfo  {journal} {Phys. Rev.}\ }\textbf {\bibinfo {volume}
  {D92}},\ \bibinfo {pages} {064041} (\bibinfo {year} {2015})},\ \Eprint
  {http://arxiv.org/abs/1507.02152} {arXiv:1507.02152 [gr-qc]} \BibitemShut
  {NoStop}%
\bibitem [{\citenamefont {{Przybilla}}\ \emph {et~al.}(2008)\citenamefont
  {{Przybilla}}, \citenamefont {{Nieva}},\ and\ \citenamefont
  {{Butler}}}]{przybilla}%
  \BibitemOpen
  \bibfield  {author} {\bibinfo {author} {\bibfnamefont {N.}~\bibnamefont
  {{Przybilla}}}, \bibinfo {author} {\bibfnamefont {M.-F.}\ \bibnamefont
  {{Nieva}}}, \ and\ \bibinfo {author} {\bibfnamefont {K.}~\bibnamefont
  {{Butler}}},\ }\href {\doibase 10.1086/595618} {\bibfield  {journal}
  {\bibinfo  {journal} {Astrophys.J.Lett.}\ }\textbf {\bibinfo {volume}
  {688}},\ \bibinfo {pages} {L103} (\bibinfo {year} {2008})},\ \Eprint
  {http://arxiv.org/abs/0809.2403} {arXiv:0809.2403} \BibitemShut {NoStop}%
\bibitem [{\citenamefont {{Bourke}}\ \emph
  {et~al.}(1995{\natexlab{a}})\citenamefont {{Bourke}}, \citenamefont
  {{Hyland}},\ and\ \citenamefont {{Robinson}}}]{1995MNRAS.276.1052B}%
  \BibitemOpen
  \bibfield  {author} {\bibinfo {author} {\bibfnamefont {T.~L.}\ \bibnamefont
  {{Bourke}}}, \bibinfo {author} {\bibfnamefont {A.~R.}\ \bibnamefont
  {{Hyland}}}, \ and\ \bibinfo {author} {\bibfnamefont {G.}~\bibnamefont
  {{Robinson}}},\ }\href@noop {} {\bibfield  {journal} {\bibinfo  {journal}
  {MNRAS}\ }\textbf {\bibinfo {volume} {276}},\ \bibinfo {pages} {1052}
  (\bibinfo {year} {1995}{\natexlab{a}})}\BibitemShut {NoStop}%
\bibitem [{\citenamefont {{Bourke}}\ \emph
  {et~al.}(1995{\natexlab{b}})\citenamefont {{Bourke}}, \citenamefont
  {{Hyland}}, \citenamefont {{Robinson}}, \citenamefont {{James}},\ and\
  \citenamefont {{Wright}}}]{1995MNRAS.276.1067B}%
  \BibitemOpen
  \bibfield  {author} {\bibinfo {author} {\bibfnamefont {T.~L.}\ \bibnamefont
  {{Bourke}}}, \bibinfo {author} {\bibfnamefont {A.~R.}\ \bibnamefont
  {{Hyland}}}, \bibinfo {author} {\bibfnamefont {G.}~\bibnamefont
  {{Robinson}}}, \bibinfo {author} {\bibfnamefont {S.~D.}\ \bibnamefont
  {{James}}}, \ and\ \bibinfo {author} {\bibfnamefont {C.~M.}\ \bibnamefont
  {{Wright}}},\ }\href@noop {} {\bibfield  {journal} {\bibinfo  {journal}
  {MNRAS}\ }\textbf {\bibinfo {volume} {276}},\ \bibinfo {pages} {1067}
  (\bibinfo {year} {1995}{\natexlab{b}})}\BibitemShut {NoStop}%
\bibitem [{\citenamefont {{Clemens}}\ and\ \citenamefont
  {{Barvainis}}(1988)}]{CB}%
  \BibitemOpen
  \bibfield  {author} {\bibinfo {author} {\bibfnamefont {D.~P.}\ \bibnamefont
  {{Clemens}}}\ and\ \bibinfo {author} {\bibfnamefont {R.}~\bibnamefont
  {{Barvainis}}},\ }\href {\doibase 10.1086/191288} {\bibfield  {journal}
  {\bibinfo  {journal} {Astrophys.J.Suppl.}\ }\textbf {\bibinfo {volume}
  {68}},\ \bibinfo {pages} {257} (\bibinfo {year} {1988})}\BibitemShut
  {NoStop}%
\bibitem [{\citenamefont {{Fleck}}(1980)}]{1980fleck}%
  \BibitemOpen
  \bibfield  {author} {\bibinfo {author} {\bibfnamefont {R.~C.}\ \bibnamefont
  {{Fleck}}, \bibfnamefont {Jr.}},\ }\href {\doibase 10.1086/158533} {\bibfield
   {journal} {\bibinfo  {journal} {Astrophys.J.}\ }\textbf {\bibinfo {volume}
  {242}},\ \bibinfo {pages} {1019} (\bibinfo {year} {1980})}\BibitemShut
  {NoStop}%
\bibitem [{\citenamefont {{Yun}}\ and\ \citenamefont
  {{Clemens}}(1990)}]{1990ApJ...365L..73Y}%
  \BibitemOpen
  \bibfield  {author} {\bibinfo {author} {\bibfnamefont {J.~L.}\ \bibnamefont
  {{Yun}}}\ and\ \bibinfo {author} {\bibfnamefont {D.~P.}\ \bibnamefont
  {{Clemens}}},\ }\href {\doibase 10.1086/185891} {\bibfield  {journal}
  {\bibinfo  {journal} {Astrophys.J.Lett.}\ }\textbf {\bibinfo {volume}
  {365}},\ \bibinfo {pages} {L73} (\bibinfo {year} {1990})}\BibitemShut
  {NoStop}%
\bibitem [{\citenamefont {{Launhardt, R.}}\ \emph {et~al.}(2013)\citenamefont
  {{Launhardt, R.}}, \citenamefont {{Stutz, A. M.}}, \citenamefont
  {{Schmiedeke, A.}}, \citenamefont {{Henning, Th.}}, \citenamefont {{Krause,
  O.}}, \citenamefont {{Balog, Z.}}, \citenamefont {{Beuther, H.}},
  \citenamefont {{Birkmann, S.}}, \citenamefont {{Hennemann, M.}},
  \citenamefont {{Kainulainen, J.}}, \citenamefont {{Khanzadyan, T.}},
  \citenamefont {{Linz, H.}}, \citenamefont {{Lippok, N.}}, \citenamefont
  {{Nielbock, M.}}, \citenamefont {{Pitann, J.}}, \citenamefont {{Ragan, S.}},
  \citenamefont {{Risacher, C.}}, \citenamefont {{Schmalzl, M.}}, \citenamefont
  {{Shirley, Y. L.}}, \citenamefont {{Stecklum, B.}}, \citenamefont
  {{Steinacker, J.}},\ and\ \citenamefont {{Tackenberg, J.}}}]{schmalzl2014}%
  \BibitemOpen
  \bibfield  {author} {\bibinfo {author} {\bibnamefont {{Launhardt, R.}}},
  \bibinfo {author} {\bibnamefont {{Stutz, A. M.}}}, \bibinfo {author}
  {\bibnamefont {{Schmiedeke, A.}}}, \bibinfo {author} {\bibnamefont {{Henning,
  Th.}}}, \bibinfo {author} {\bibnamefont {{Krause, O.}}}, \bibinfo {author}
  {\bibnamefont {{Balog, Z.}}}, \bibinfo {author} {\bibnamefont {{Beuther,
  H.}}}, \bibinfo {author} {\bibnamefont {{Birkmann, S.}}}, \bibinfo {author}
  {\bibnamefont {{Hennemann, M.}}}, \bibinfo {author} {\bibnamefont
  {{Kainulainen, J.}}}, \bibinfo {author} {\bibnamefont {{Khanzadyan, T.}}},
  \bibinfo {author} {\bibnamefont {{Linz, H.}}}, \bibinfo {author}
  {\bibnamefont {{Lippok, N.}}}, \bibinfo {author} {\bibnamefont {{Nielbock,
  M.}}}, \bibinfo {author} {\bibnamefont {{Pitann, J.}}}, \bibinfo {author}
  {\bibnamefont {{Ragan, S.}}}, \bibinfo {author} {\bibnamefont {{Risacher,
  C.}}}, \bibinfo {author} {\bibnamefont {{Schmalzl, M.}}}, \bibinfo {author}
  {\bibnamefont {{Shirley, Y. L.}}}, \bibinfo {author} {\bibnamefont
  {{Stecklum, B.}}}, \bibinfo {author} {\bibnamefont {{Steinacker, J.}}}, \
  and\ \bibinfo {author} {\bibnamefont {{Tackenberg, J.}}},\ }\href {\doibase
  10.1051/0004-6361/201220477} {\bibfield  {journal} {\bibinfo  {journal}
  {A\&A}\ }\textbf {\bibinfo {volume} {551}},\ \bibinfo {pages} {A98} (\bibinfo
  {year} {2013})}\BibitemShut {NoStop}%
\bibitem [{\citenamefont {{Nelson}}\ and\ \citenamefont
  {{Langer}}(1999)}]{1999ApJ...524..923N}%
  \BibitemOpen
  \bibfield  {author} {\bibinfo {author} {\bibfnamefont {R.~P.}\ \bibnamefont
  {{Nelson}}}\ and\ \bibinfo {author} {\bibfnamefont {W.~D.}\ \bibnamefont
  {{Langer}}},\ }\href {\doibase 10.1086/307823} {\bibfield  {journal}
  {\bibinfo  {journal} {Astrophys.J.}\ }\textbf {\bibinfo {volume} {524}},\
  \bibinfo {pages} {923} (\bibinfo {year} {1999})}\BibitemShut {NoStop}%
\bibitem [{\citenamefont {{Hartley}}\ \emph {et~al.}(1986)\citenamefont
  {{Hartley}}, \citenamefont {{Tritton}}, \citenamefont {{Manchester}},
  \citenamefont {{Smith}},\ and\ \citenamefont {{Goss}}}]{1986hartley}%
  \BibitemOpen
  \bibfield  {author} {\bibinfo {author} {\bibfnamefont {M.}~\bibnamefont
  {{Hartley}}}, \bibinfo {author} {\bibfnamefont {S.~B.}\ \bibnamefont
  {{Tritton}}}, \bibinfo {author} {\bibfnamefont {R.~N.}\ \bibnamefont
  {{Manchester}}}, \bibinfo {author} {\bibfnamefont {R.~M.}\ \bibnamefont
  {{Smith}}}, \ and\ \bibinfo {author} {\bibfnamefont {W.~M.}\ \bibnamefont
  {{Goss}}},\ }\href@noop {} {\bibfield  {journal} {\bibinfo  {journal}
  {Astron. Astrophys. Suppl. Ser.}\ }\textbf {\bibinfo {volume} {63}},\
  \bibinfo {pages} {27} (\bibinfo {year} {1986})}\BibitemShut {NoStop}%
\bibitem [{\citenamefont {{Kandori}}\ \emph {et~al.}(2005)\citenamefont
  {{Kandori}}, \citenamefont {{Nakajima}}, \citenamefont {{Tamura}},
  \citenamefont {{Tatematsu}}, \citenamefont {{Aikawa}}, \citenamefont
  {{Naoi}}, \citenamefont {{Sugitani}}, \citenamefont {{Nakaya}}, \citenamefont
  {{Nagayama}}, \citenamefont {{Nagata}}, \citenamefont {{Kurita}},
  \citenamefont {{Kato}}, \citenamefont {{Nagashima}},\ and\ \citenamefont
  {{Sato}}}]{2005Kandori}%
  \BibitemOpen
  \bibfield  {author} {\bibinfo {author} {\bibfnamefont {R.}~\bibnamefont
  {{Kandori}}}, \bibinfo {author} {\bibfnamefont {Y.}~\bibnamefont
  {{Nakajima}}}, \bibinfo {author} {\bibfnamefont {M.}~\bibnamefont
  {{Tamura}}}, \bibinfo {author} {\bibfnamefont {K.}~\bibnamefont
  {{Tatematsu}}}, \bibinfo {author} {\bibfnamefont {Y.}~\bibnamefont
  {{Aikawa}}}, \bibinfo {author} {\bibfnamefont {T.}~\bibnamefont {{Naoi}}},
  \bibinfo {author} {\bibfnamefont {K.}~\bibnamefont {{Sugitani}}}, \bibinfo
  {author} {\bibfnamefont {H.}~\bibnamefont {{Nakaya}}}, \bibinfo {author}
  {\bibfnamefont {T.}~\bibnamefont {{Nagayama}}}, \bibinfo {author}
  {\bibfnamefont {T.}~\bibnamefont {{Nagata}}}, \bibinfo {author}
  {\bibfnamefont {M.}~\bibnamefont {{Kurita}}}, \bibinfo {author}
  {\bibfnamefont {D.}~\bibnamefont {{Kato}}}, \bibinfo {author} {\bibfnamefont
  {C.}~\bibnamefont {{Nagashima}}}, \ and\ \bibinfo {author} {\bibfnamefont
  {S.}~\bibnamefont {{Sato}}},\ }\href {\doibase 10.1086/444619} {\bibfield
  {journal} {\bibinfo  {journal} {Astron.J.}\ }\textbf {\bibinfo {volume}
  {130}},\ \bibinfo {pages} {2166} (\bibinfo {year} {2005})},\ \Eprint
  {http://arxiv.org/abs/astro-ph/0506205} {astro-ph/0506205} \BibitemShut
  {NoStop}%
\bibitem [{\citenamefont {De~Felice}\ and\ \citenamefont
  {Tsujikawa}(2010)}]{DeFelice:2010aj}%
  \BibitemOpen
  \bibfield  {author} {\bibinfo {author} {\bibfnamefont {A.}~\bibnamefont
  {De~Felice}}\ and\ \bibinfo {author} {\bibfnamefont {S.}~\bibnamefont
  {Tsujikawa}},\ }\href@noop {} {\bibfield  {journal} {\bibinfo  {journal}
  {Living Rev. Rel.}\ }\textbf {\bibinfo {volume} {13}},\ \bibinfo {pages} {3}
  (\bibinfo {year} {2010})},\ \Eprint {http://arxiv.org/abs/1002.4928}
  {arXiv:1002.4928 [gr-qc]} \BibitemShut {NoStop}%
\bibitem [{\citenamefont {Bunn}\ and\ \citenamefont {White}(1997)}]{Bunn1996}%
  \BibitemOpen
  \bibfield  {author} {\bibinfo {author} {\bibfnamefont {E.~F.}\ \bibnamefont
  {Bunn}}\ and\ \bibinfo {author} {\bibfnamefont {M.~J.}\ \bibnamefont
  {White}},\ }\href {\doibase 10.1086/303955} {\bibfield  {journal} {\bibinfo
  {journal} {Astrophys. J.}\ }\textbf {\bibinfo {volume} {480}},\ \bibinfo
  {pages} {6} (\bibinfo {year} {1997})},\ \Eprint
  {http://arxiv.org/abs/astro-ph/9607060} {arXiv:astro-ph/9607060} \BibitemShut
  {NoStop}%
\bibitem [{\citenamefont {Schmidt}\ \emph {et~al.}(2009)\citenamefont
  {Schmidt}, \citenamefont {Lima}, \citenamefont {Oyaizu},\ and\ \citenamefont
  {Hu}}]{Schmidt:2008tn}%
  \BibitemOpen
  \bibfield  {author} {\bibinfo {author} {\bibfnamefont {F.}~\bibnamefont
  {Schmidt}}, \bibinfo {author} {\bibfnamefont {M.~V.}\ \bibnamefont {Lima}},
  \bibinfo {author} {\bibfnamefont {H.}~\bibnamefont {Oyaizu}}, \ and\ \bibinfo
  {author} {\bibfnamefont {W.}~\bibnamefont {Hu}},\ }\href {\doibase
  10.1103/PhysRevD.79.083518} {\bibfield  {journal} {\bibinfo  {journal} {Phys.
  Rev.}\ }\textbf {\bibinfo {volume} {D79}},\ \bibinfo {pages} {083518}
  (\bibinfo {year} {2009})},\ \Eprint {http://arxiv.org/abs/0812.0545}
  {arXiv:0812.0545 [astro-ph]} \BibitemShut {NoStop}%
\bibitem [{\citenamefont {Vikram}\ \emph {et~al.}(2013)\citenamefont {Vikram},
  \citenamefont {Cabre}, \citenamefont {Jain},\ and\ \citenamefont
  {VanderPlas}}]{Vikram:2013uba}%
  \BibitemOpen
  \bibfield  {author} {\bibinfo {author} {\bibfnamefont {V.}~\bibnamefont
  {Vikram}}, \bibinfo {author} {\bibfnamefont {A.}~\bibnamefont {Cabre}},
  \bibinfo {author} {\bibfnamefont {B.}~\bibnamefont {Jain}}, \ and\ \bibinfo
  {author} {\bibfnamefont {J.~T.}\ \bibnamefont {VanderPlas}},\ }\href
  {\doibase 10.1088/1475-7516/2013/08/020} {\bibfield  {journal} {\bibinfo
  {journal} {JCAP}\ }\textbf {\bibinfo {volume} {1308}},\ \bibinfo {pages}
  {020} (\bibinfo {year} {2013})},\ \Eprint {http://arxiv.org/abs/1303.0295}
  {arXiv:1303.0295 [astro-ph.CO]} \BibitemShut {NoStop}%
\bibitem [{\citenamefont {Jain}\ and\ \citenamefont
  {VanderPlas}(2011)}]{Jain:2011ji}%
  \BibitemOpen
  \bibfield  {author} {\bibinfo {author} {\bibfnamefont {B.}~\bibnamefont
  {Jain}}\ and\ \bibinfo {author} {\bibfnamefont {J.}~\bibnamefont
  {VanderPlas}},\ }\href {\doibase 10.1088/1475-7516/2011/10/032} {\bibfield
  {journal} {\bibinfo  {journal} {JCAP}\ }\textbf {\bibinfo {volume} {1110}},\
  \bibinfo {pages} {032} (\bibinfo {year} {2011})},\ \Eprint
  {http://arxiv.org/abs/1106.0065} {arXiv:1106.0065 [astro-ph.CO]} \BibitemShut
  {NoStop}%
\bibitem [{\citenamefont {Hui}\ \emph {et~al.}(2009)\citenamefont {Hui},
  \citenamefont {Nicolis},\ and\ \citenamefont {Stubbs}}]{Hui:2009kc}%
  \BibitemOpen
  \bibfield  {author} {\bibinfo {author} {\bibfnamefont {L.}~\bibnamefont
  {Hui}}, \bibinfo {author} {\bibfnamefont {A.}~\bibnamefont {Nicolis}}, \ and\
  \bibinfo {author} {\bibfnamefont {C.}~\bibnamefont {Stubbs}},\ }\href
  {\doibase 10.1103/PhysRevD.80.104002} {\bibfield  {journal} {\bibinfo
  {journal} {Phys. Rev.}\ }\textbf {\bibinfo {volume} {D80}},\ \bibinfo {pages}
  {104002} (\bibinfo {year} {2009})},\ \Eprint {http://arxiv.org/abs/0905.2966}
  {arXiv:0905.2966 [astro-ph.CO]} \BibitemShut {NoStop}%
\bibitem [{\citenamefont {Okada}\ \emph {et~al.}(2013)\citenamefont {Okada},
  \citenamefont {Totani},\ and\ \citenamefont {Tsujikawa}}]{Okada:2012mn}%
  \BibitemOpen
  \bibfield  {author} {\bibinfo {author} {\bibfnamefont {H.}~\bibnamefont
  {Okada}}, \bibinfo {author} {\bibfnamefont {T.}~\bibnamefont {Totani}}, \
  and\ \bibinfo {author} {\bibfnamefont {S.}~\bibnamefont {Tsujikawa}},\ }\href
  {\doibase 10.1103/PhysRevD.87.103002} {\bibfield  {journal} {\bibinfo
  {journal} {Phys.Rev.}\ }\textbf {\bibinfo {volume} {D87}},\ \bibinfo {pages}
  {103002} (\bibinfo {year} {2013})},\ \Eprint {http://arxiv.org/abs/1208.4681}
  {arXiv:1208.4681 [astro-ph.CO]} \BibitemShut {NoStop}%
\bibitem [{\citenamefont {Capozziello}\ and\ \citenamefont
  {Tsujikawa}(2008)}]{Capozziello:2007eu}%
  \BibitemOpen
  \bibfield  {author} {\bibinfo {author} {\bibfnamefont {S.}~\bibnamefont
  {Capozziello}}\ and\ \bibinfo {author} {\bibfnamefont {S.}~\bibnamefont
  {Tsujikawa}},\ }\href {\doibase 10.1103/PhysRevD.77.107501} {\bibfield
  {journal} {\bibinfo  {journal} {Phys. Rev.}\ }\textbf {\bibinfo {volume}
  {D77}},\ \bibinfo {pages} {107501} (\bibinfo {year} {2008})},\ \Eprint
  {http://arxiv.org/abs/0712.2268} {arXiv:0712.2268 [gr-qc]} \BibitemShut
  {NoStop}%
\bibitem [{\citenamefont {Roshan}\ and\ \citenamefont
  {Abbassi}(2015)}]{Roshan:2015gra}%
  \BibitemOpen
  \bibfield  {author} {\bibinfo {author} {\bibfnamefont {M.}~\bibnamefont
  {Roshan}}\ and\ \bibinfo {author} {\bibfnamefont {S.}~\bibnamefont
  {Abbassi}},\ }\href {\doibase 10.1088/0004-637X/802/1/9} {\bibfield
  {journal} {\bibinfo  {journal} {Astrophys. J.}\ }\textbf {\bibinfo {volume}
  {802}},\ \bibinfo {pages} {9} (\bibinfo {year} {2015})},\ \Eprint
  {http://arxiv.org/abs/1501.04715} {arXiv:1501.04715 [astro-ph.GA]}
  \BibitemShut {NoStop}%
\end{thebibliography}%

\end{document}